\documentclass[aps,amsfonts,prd,showpacs,nobibnotes,nofootinbib,tightenlines,floatfix,eqsecnum]{revtex4}
\usepackage{graphicx}
\usepackage{bbm}	
\usepackage{latexsym}	
\usepackage{mathrsfs}	
\begin{document}
\def\figwidth{7.5cm}

\def\cD{{\cal D}}
\def\cF{{\cal F}}
\def\cE{{\cal E}}
\def\bfn{{\bf n}}
\newcommand{\tr}{\mbox{tr}}

\newcommand{\Tr}{\mbox{Tr}}

\newcommand{\eq}[1]{eq.~(\ref{#1})}

\newcommand{\be}{\begin{equation}}

\newcommand{\ee}{\end{equation}}

\newcommand{\bea}{\begin{eqnarray}}

\newcommand{\eea}{\end{eqnarray}}

\newcommand{\zr}[1]{\mbox{\hspace*{#1em}}}

\newcommand{\ID}{\mbox{$\mathbbm{1}$}}

\newcommand{\oneF}{_{\rm eff}^{(1)}}

\newcommand{\NN}{\mathscr{N}}
\newcommand{\MM}{\mathscr{M}}
\newcommand{\OO}{\mathscr{O}}
\newcommand{\DD}{\mathscr{D}}
\newcommand{\GG}{\mathscr{G}}
\newcommand{\HH}{\mathscr{H}}
\newcommand{\FF}{\mathscr{F}}
\newcommand{\YY}{\mathscr{Y}}
\newcommand{\LL}{\mathscr{L}}
\newcommand{\Sc}{\mathscr{S}}
\newcommand{\vek}[1]{\mathbf{#1}}

\newcommand{\HiggsGauge}{Higgs--gauge~}

\newcommand{\azAngle}{\varphi}

\newcommand{\half}{\frac{1}{2}}

\newcommand{\vev}{VEV}
%
%
\def\Ord#1{{\cal O}\left( #1\right)}
\renewcommand{\Im}{{\rm Im}}

%

%
%


\title{Quantum Energies of Strings in a
2+1 Dimensional Gauge Theory}
\author{N.~Graham} \email{ngraham@middlebury.edu}
\affiliation{Department of Physics, Middlebury College \\
Middlebury, VT 05753, USA}
\author{M.~Quandt} \email{quandt@tphys.physik.uni-tuebingen.de}
\affiliation{Institute for Theoretical Physics, T\"ubingen University\\
D-72076 T\"ubingen, Germany}
\author{O.~Schr\"oder} \email{oliver.schroeder@science-computing.de}
\affiliation{science+computing ag\\Hagellocher Weg 73, 72070
  T\"ubingen, FRG}
\author{H.~Weigel} \email{weigel@physik.uni-siegen.de}
\affiliation{Fachbereich Physik, Siegen University\\
D-57068 Siegen, Germany}
\begin{abstract}
We study classically unstable string type configurations and compute
the renormalized vacuum polarization energies that arise from fermion
fluctuations in a 2+1 dimensional analog of the standard model. 
We then search for a minimum of the total energy (classical plus vacuum 
polarization energies) by varying the profile functions that
characterize the string. We find that typical string configurations
bind numerous fermions and that populating these levels is
beneficial to further decrease the total energy. Ultimately our goal
is to explore the stabilization of string type
configurations in the standard model through quantum effects.

We compute the vacuum polarization energy within the phase shift formalism 
which identifies terms in the Born series for scattering data and Feynman
diagrams. This approach allows us to implement standard renormalization 
conditions of perturbation theory and thus yields the unambiguous result for
this non--perturbative contribution to the total energy.
\end{abstract}
\vspace*{-\bigskipamount}
\maketitle

\section{Introduction}

In the past decades the perturbative treatment of the electroweak 
standard model has proven to be a very powerful tool to describe the 
properties and interactions of elementary particles within a wide 
energy regime. On the other hand the role and even the existence of 
non--perturbative solutions in this model is still quite uncertain.
While the standard model does not contain topological
solitons, one can construct nontopological string solutions to the
classical equations of motion, called $Z$-strings or electroweak strings
\cite{Vachaspati:1992fi,VachaspatiReview, Nambu}. In the absence of 
topological arguments, however, one is not guaranteed that these
classical solutions actually correspond to true local minima of the
full effective energy, or even the classical energy.  Indeed, 
Naculich~\cite{Naculich:1995cb} has shown that in the limit of weak 
coupling, fermion fluctuations destabilize the string solution.  
This analysis leaves
open the possibility, however, that deformed string solutions could
exist, which would be local minima of the full effective energy.
Since fermions are tightly bound in the string background, one
potential mechanism for restoring stability is for fermions to bind
to the string, yielding a lower total energy than a corresponding
density of free fermions.  
When including this effect, however, one must also
take into account the shifts in the zero-point energies of all the
unoccupied modes as well, computed consistently in a standard
renormalization scheme.

If stabilized by the fermion binding mechanism,  electroweak strings could
have significant cosmological consequences \cite{Kibble,Hindmarsh:1994re}.  
A network of strings could contribute to the dark energy that is required 
to explain the recently observed cosmic 
acceleration~\cite{Perlmutter:1998np,Riess:1998cb}. However, 
a complete dynamical description of that scenario is still 
missing~\cite{Spergel:1996ai,McGraw:1997nx,Bucher:1998mh}.
Also, as pointed out by Nambu~\cite{Nambu}, Z--strings are expected 
to terminate in monopole--antimonopole pairs, which could give rise 
to a primordial magnetic field.  Furthermore, a network of stable 
strings at the electroweak phase transition would provide a scenario 
for electroweak baryogenesis without requiring a first--order phase
transition~\cite{Brandenberger}. The strings provide out--of--equilibrium 
regions, and the core of the string has copious baryon number violation 
due to the suppressed Higgs condensate. They thus provide an alternative 
to the usual idea of bubble--nucleation baryogenesis, which requires a 
first order phase transition to go out of thermal equilibrium.  

From a theoretical point of view, the quantum properties of
$Z$-strings have been connected to non-perturbative anomalies
\cite{Klinkhamer:2003hz}.  Decoupling arguments 
\cite{D'Hoker:1984ph,D'Hoker:1984pc} suggest
that when a fermion's mass is made very large by increasing its Yukawa
coupling, soliton configurations should appear in the low-energy
spectrum to maintain cancellation of such anomalies.  In these
calculations, the analysis of the fermion determinant -- whose logarithm
is the sum over zero-point energies -- is essential.

A first attempt at a full calculation of the quantum corrections to
the $Z$-string energy was carried out in \cite{Groves:1999ks}.
Those authors were only able to compare the energies of two
string configurations, rather than comparing a single string
configuration to the vacuum. Furthermore, they used a proper 
time prescription, in which it was not possible to explicitly 
split off the divergent parts from the fermion determinant. Thus
 the final answer was expressed at large but finite cut--off in terms 
of two large terms that could only be computed numerically, representing 
the fermion determinant and the counterterms. In this formulation, no
numerically stable results could be extracted for the limit 
``cut--off to infinity.'' In the present work, we will employ phase 
shift techniques \cite{Leipzig}, which allow us to make this separation 
cleanly and unambiguously.

In this paper we restrict ourselves to the case of $2+1$ dimensions.
We make this restriction to simplify the renormalization
procedure. However, from the experience gathered in the case of QED
magnetic flux tubes~\cite{Graham:2004jb}, it is reasonable to expect that
the qualitative behavior might be very similar to the actual 3+1
dimensional theory.  The techniques of \cite{GrahamInterface} allow
for a straightforward generalization of the calculational procedure
used here to that case as well.  We also consider only the weak
interactions, neglecting electromagnetism.  Although introducing
electromagnetism into our calculation is not entirely 
straightforward because of the effects of Aharonov-Bohm phases discussed in
ref.~\cite{Graham:2004jb}, we expect the approach of that paper to provide
a natural generalization to the full electroweak interactions.  In
particular, the symmetry operator we rely on for the partial wave
decomposition of our phase shifts continues to hold in the full theory.

The paper is organized as follows. In Section II we will
consider the Higgs--gauge sector model, consisting of a
doublet Higgs coupled to an $SU(2)$ gauge field, and discuss various
aspects of classical string configurations. In Section III we discuss
the coupling of the fermions to string configurations and compute the
energy that arises from summing the fermion zero modes. We will
present our numerical results for the corrected string energy in 
Section IV. Some preliminary results of the current investigation
were reported in conference proceedings~\cite{Schroeder:2006hk}.

\section{The Higgs--Gauge--Sector}
\noindent
Our principal aim is to study string-like configurations within the
standard electroweak $SU(2) \times U(1)$ model. In the present paper, we
will consider a simplified version of this theory, in which
\begin{enumerate}
\item the spacetime dimension is reduced to $D=2+1$ or, equivalently,
bosonic configurations are translationally invariant in the third 
space direction,
\item the Weinberg angle, $\theta_{\rm W}$ is set to zero, 
{\it i.e.\@}~the $U(1)$ factor in the gauge group decouples (and 
will be discarded in the following). The classical energy actually 
decreases when $\theta_{\rm W}$ is turned on in the full theory. 
Hence a configuration that is stable at $\theta_{\rm W}=0$ most likely is 
also stable otherwise. And
\item quantum corrections to the bosonic string configurations are
taken from the fermion sector only; this approximation may be justified
in situations where the degeneracy of the fermion modes (e.g.~the number
of colors $N_C$) is large.
\end{enumerate}
The action of the simplified model consists of two parts,
$S = S_H + S_F$, where the bosonic (Higgs-gauge) sector $S_H$ is treated
\emph{classically} and the fermion part gives rise to quantum fluctuations
around bosonic field configurations. In the present section, we
discuss properties of classical string-like configurations in the
bosonic sector while the quantum treatment of the fermions is deferred
to section \ref{sec:FermionsOnStrings}.

The Higgs-gauge sector $S_H$ in our model reads
\bea
S_H [ \phi, W] & = & \int d^3 x \left[ -\frac{1}{2} \tr
  \left(G^{\mu\nu}G_{\mu\nu}\right) +
\left(D^{\mu}\phi \right)^{\dag} D_{\mu}\phi - \lambda \left(
     \phi^{\dag} \phi - v^2 \right)^2 \right] \, .
\label{eq:HiggsLagrangian}
\eea
We work in $D=2+1$ spacetime with Minkowski signature and include the gauge
coupling strength, $g$, explicitly in all vertices. Thus, the field strength
tensor is given by
\be
G_{\mu \nu} = \partial_\mu W_\nu - \partial_\nu W_\mu - i g \,[ W_\mu, W_\nu ]
\, ,
\ee
where the three $SU(2)$ vector bosons, $W_\mu^a$, are components of a 
Lie--algebra valued matrix connection,
\be
W_\mu = W_\mu^a \,\frac{\tau^a}{2} \equiv W_\mu^a\,T^a \, .
\ee
As indicated, we use hermitian generators $T^a = \tau^a/2$
(where $\tau^a$ are the Pauli matrices) with the commutation relation
$[T^a,T^b] =i \epsilon^{abc}\,T^c$. The complex scalar Higgs field is
in the fundamental representation of $SU(2)$,
\be
\phi =  \left( \phi_+ \atop \phi_0 \right)
\ee
while the covariant derivative contains the Higgs--gauge coupling constant
$g$ explicitly,
\be
D_\mu \phi = \left( \partial_\mu  - i g \,W_\mu \right) \phi \, .
  \label{eq:covariantDerivative}
\ee
By construction, the action eq.~(\ref{eq:HiggsLagrangian}) is invariant
under $SU(2)$ gauge rotations  $V=\exp(i\theta^a(x) T^a)$,
\be
\phi \rightarrow  V \phi \qquad {\rm and}\qquad
ig\,W_\mu \rightarrow V \left(-\partial_\mu + i g\, W_\mu \right) V^\dag \, .
\label{eq:GaugeTransform}
\ee

\noindent
For positive values of the self-coupling $\lambda>0$, the Higgs doublet
acquires a vacuum expectation value (\vev) conventionally chosen
as\footnote{In the full $SU(2)\times U(1)$ model, this choice ensures that the
unbroken $U(1)$ group is generated by the electric charge $Q = I_w^3 + Y_W/2$,
where $I_w$ and $Y_W$ are weak isospin and hypercharge, respectively.}
\[
\langle \phi_0 \rangle = v\,,\qquad\quad \langle \phi_+ \rangle = 0\,.
\]
As a consequence, the system undergoes complete spontaneous symmetry
breaking $SU(2) \to \{ \mathbbm{1} \}$, i.e.~there is no group generator that
leaves the vacuum invariant. As a consequence, there are three
would-be Goldstone modes from $\phi$ fluctuations, which become
longitudinal modes of the gauge field as it acquires a mass
$M_{\rm W}=gv/\sqrt{2}$ (at tree level). The remaining Higgs excitations
are also massive with a (tree-level) mass of $M_{\rm H}=2v\sqrt{\lambda}$.

In the course of this paper it will occasionally be convenient to
re-write the covariant derivative, eq.~(\ref{eq:covariantDerivative}) as
\be
D_\mu \phi=\left[\partial_\mu-ig Q_Z Z_\mu-i g
\left(W_\mu^+T^-+W_\mu^-T^+\right)\right] \phi \, ,
\ee
where the charge operators and fields are 
\be
\begin{array}{r@{\,\,=\,\,}l@{\qquad\qquad}r@{\,\,=\,\,}l}
Q_Z & \displaystyle \half\tau^3 &
T^\pm &  \displaystyle \frac{1}{\sqrt{2}}\left(\tau^1/2\pm
i\,\tau^2/2\right)
\\[4mm]
Z_\mu &  W_\mu^3 & W_\mu^\pm &
\displaystyle \frac{1}{\sqrt{2}}\left(W_\mu^1\pm i W_\mu^2\right)\,.
\end{array}
\label{eq:charges}
\ee
We will use the notation $W^3_\mu$ and $Z_\mu$ interchangeably.

\subsection{The Classical String Solutions}

Our main interest is in string-like configurations of the bosonic 
sector.  We analyze the energy of (classical) localized configurations
in two space dimensions. A subsequent extension of this analysis
would be to consider configurations in three dimensions that are
translationally invariant in the $z$ direction, with a finite energy
per unit length in a localized region of the $x-y$ plane. The region
in which the energy density is concentrated would thus take the form
of a tube extending along the z-axis.  Such field configurations are
commonly called \emph{cosmic strings} to distinguish them from the
fundamental objects in string theory. In the present paper, we will
refer to these objects simply as \emph{strings} or \emph{vortices}.

Finite energy requires the configuration to be pure gauge at spatial
infinity. Let $(\rho,\azAngle)$ denote the radial and angular
coordinates in the plane. At $\rho \to \infty$, the most general
vacuum configuration is
\be
\phi^{(\infty)}=V\: \left( 0 \atop v \right)
\qquad {\rm and} \qquad
g W_\mu^{(\infty)}= iV\partial_\mu V^\dag \, ,
\label{eq:vacconfig}
\ee
where the $SU(2)$ rotation $V(\azAngle)$ maps the circle $S_1$ at
infinity onto the gauge group $G$. For the electromagnetic case
$G= U(1)$, such maps would be characterized by an integral winding
number $\Pi_1 (U(1)) = \mathbbm{Z}$ and we would obtain topologically stable
Nielsen--Olesen vortices \cite{Nielsen:1973cs}. However, the weak
isospin group in our model is $G=SU(2)$. We can still embed
Nielsen--Olesen type solutions in this larger gauge group by
considering maps from the spatial boundary $S_1$ to various
$U(1)$ subgroups of $G=SU(2)$. There is, of course, considerable
arbitrariness in this procedure, since no $U(1)$ subgroup survives
the electroweak symmetry breaking, so there is no 'natural' 
candidate for the target $U(1)$. Moreover, from $\Pi_1(SU(2)) = 0$, no 
such embedded string configuration is \emph{topologically stable} 
against deformations in the full $SU(2)$ group; this will be
discussed in more detail in section \ref{sec:interpolation}.

First we consider the $U(1)$ subgroup generated by the charge $Q_Z$ associated
with the neutral $W^3_\mu$ (or $Z_\mu$) gauge fields, as defined in
eq.~(\ref{eq:charges}). In the full $SU(2)\times U(1)$ electroweak theory,
$Q_Z$ becomes the charge of $Z_\mu$, the gauge field that couples 
to the weak neutral current, and the vortex associated with $Q_Z$
becomes the \emph{Z-string}. The maps that specify the asymptotic pure-gauge
configurations are
\be
V(\azAngle)=e^{-2in\azAngle\,Q_Z} \, ,
\ee
where $n$ must be integral to obtain single-valued Higgs fields.
To construct the corresponding vortex configuration, we first adopt the Weyl
gauge
\[
W_0^a = 0\,.
\]
In non-Abelian gauge theories, there might be topological obstructions to
implementing this gauge \cite{Reinhardt:1997b, Quandt:1999a}; such defects are,
however, associated with non-trivial (periodic) boundary conditions in the
\emph{time} direction. For static configurations, no such obstructions 
exists and the Weyl gauge may always be attained; the residual gauge freedom
consists of all time-independent gauge rotations. We shall adopt this
gauge throughout the rest of the paper.

The explicit form of the $W^3/Z$-string now dwells in the neutral
component of the gauge field ($\hat{\azAngle}$ is the unit vector in
azimuthal direction)
\be
\phi_0=vf_H(\rho)\,{\rm e}^{i n \azAngle} \qquad {\rm and}\qquad
\vec{W}^3=\frac{2 n}{g \rho}\,f_G(\rho)\hat{\azAngle} \, ,
\label{eq:w3zstring}
\ee
with all other fields vanishing. The radial functions $f_H$ and $f_G$ must
vanish at the origin to avoid singularities, and approach unity at infinity
to have finite energy. Substituting this \emph{ansatz} into the action,
eq.~(\ref{eq:HiggsLagrangian}), yields the classical
energy
\be
E_{\rm cl} = 2\pi \int_0^\infty d\rho \, \rho \Biggl[
\frac{2}{g^2} \left( \frac{f_G'}{\rho}  \right)^2 + v^2 \left( f_H'
\right)^2 + \frac{v^2}{\rho^2} f_H^2 \left( 1 - f_G \right)^2
 + \lambda v^4 \left(1 - f_H^2 \right)^2 \Biggr] \, .
\label{eq:eclslsq}
\ee
Minimizing this energy functional leads to the Nielsen--Olesen
differential equations
\bea
0 & = & f_H'' + \frac{1}{\rho} f_H' - \frac{n^2}{\rho^2}f_H ( 1 - f_G )^2 + 2
\lambda v^2 f_H (1 - f_H^2) \, , \nonumber \\
0 & = & f_G'' - \frac{1}{\rho} f_G' + \half (g^2+g'^2) v^2 f_H^2 (1-f_G) \, ,
\label{eq:NOequations}
\eea
which can be integrated numerically. Near the origin $\rho\to 0$, the radial
functions behave as
\be
f_H(\rho)  \propto \rho \qquad {\rm and} \qquad
f_G(\rho)  \propto  \rho^2 \, .
\ee
Notice finally that our $W^3/Z$-vortex carries quantized magnetic flux
associated with the Abelian charge $Q_Z$. Defining the pseudoscalar
magnetic field $B=\partial_x W^3_y-\partial_y W^3_x$, we have
\be
F\,=\,\int d^2x\,B\,=\,\frac{4\pi n}{g}\, .
\ee
This flux would flow along the vortex tube in $D=3+1$ dimensions.
The classical energy density creates a pressure that tends to spread
out the flux, while the Higgs condensate is suppressed in the region
of non-vanishing $B$, which tends to compress the flux. As a result of
these two competing effects, the $W^3/Z$-string is stabilized at an
equilibrium thickness.

In the above exploration we have chosen to embed the string in the 
$W^3/Z$--$U(1)$ submanifold. This choice, however, is arbitrary. For 
example, we could equally well chose a family of $U(1)$ submanifolds that
is defined by the generator $\hat{n}\cdot\vec{T}$ where $\hat{n}$ 
is a unit vector.

In addition to strings carried by the neutral gauge field, there are also
string-like configurations that live entirely in the \emph{charged} sector
with $Z_\mu\equiv0$.

\subsection{Connecting Strings to Vacua: The Sphaleron Square}

\label{sec:interpolation}
String solutions are characterized by the homotopy group $\Pi_1(G)$ of their
boundary condition (\ref{eq:vacconfig}) at spatial infinity. Since
$\Pi_1(SU(2)) = 0$, no topologically stable string solutions exists in our
model (nor in the full electroweak theory). In other words, the string
solution can be continuously deformed into the vacuum with finite
energies along the entire path of deformation.  In the following we
will consider such deformation paths, because they represent the
starting point for a variational approach to minimize the string
energy when the vacuum polarization from the fermion sector is included.

\subsubsection{String deformations}

We construct non--contractible two--parameter loops
of configurations with W-strings as the top of the tightest
loop~\cite{Klinkhamer:1994uy,KhemaniThesis}: Let $\beta_1,
\beta_2, \alpha$ denote angular coordinates which describe a
three-dimensional sphere, $S_3$, with $0 \le \beta_i \le \pi$ and $0 \le
\alpha < 2\pi$. The sphere $S_3$ can be embedded as a unit sphere in a
four-dimensional Euclidean space and is described by unit vectors
\be
\hat{n} (\beta_1, \beta_2, \alpha) =
\pmatrix{
  \sin\beta_1 \sin\beta_2 \cos\alpha \cr
  \cos \beta_1 \cr
  \sin\beta_1 \cos\beta_2 \cr
  \sin\beta_1 \sin\beta_2 \sin\alpha}\,.
\ee
First, a map with unit winding number from $S_3$ to $SU(2)$ is given by
\be
U^{(1)} (\beta_1, \beta_2, \alpha) = \hat{n}_0 \ID -
i \hat{n} \cdot  {\vec\tau} \, .
\label{eq:S3_1}
\ee
To construct a map with winding number $n$, we may either raise the above
$U^{(1)}$ to the power of $n$ or, even simpler, scale the azimuthal angle
$\alpha$ of $U^{(1)}$~\cite{Weigel:1986zc},
\be
U^{(n)} (\beta_1, \beta_2, \alpha) = U^{(1)} (\beta_1, \beta_2, n \alpha) \, .
\ee
If we now identify the $S_1$ subspace spanned by $\alpha$ with the circle at
infinity in the spatial $x-y$ plane ({\it i.e.\@} we set $\azAngle = \alpha$),
we obtain a two--parameter family of maps characterizing possible
string boundary conditions
(\ref{eq:vacconfig}):
\be
U_{\xi_1, \xi_2}(\azAngle) = U^{(n)}(\xi_1, \xi_2, \azAngle) \, .
\ee
As before, we construct string configurations with these boundary conditions
by dressing the asymptotic form, eq.~(\ref{eq:vacconfig}), with appropriate
radial functions,
\be
\phi = f_H(\rho) U_{\xi_1, \xi_2}(\azAngle) \left( 0 \atop v \right)
\qquad {\rm and} \qquad
\vec{W} = \frac{1}{g}\, \frac{\hat{\azAngle}}{\rho} f_G(\rho)
U_{\xi_1,\xi_2}(\azAngle) \partial_\azAngle U_{\xi_1, \xi_2}^\dag(\azAngle) \,,
\ee
The radial functions go from zero at the origin (to ensure smoothness)
and to unity at infinity (to ensure finite energy).
Substituting the explicit expression for $U_{\xi_1, \xi_2}$, we get
\bea
\phi & = & v f_H(\rho)\, \left(\begin{array}{c}
\displaystyle -i \cos \xi_1 - \sin \xi_1 \cos \xi_2 \\[1mm]
\displaystyle\sin \xi_1 \sin \xi_2  {\rm e}^{i n \azAngle} \end{array}\right)
\nonumber \\[2mm]
\vec{W}^3&=& \frac{2n}{g \rho}\,\hat{\azAngle} f_G(\rho)\, \sin^2 \xi_1
\,\sin^2 \xi_2 \, , \nonumber \\[2mm]
\vec{W}^+ &=& \frac{\sqrt{2}n}{g \rho}\,\hat{\azAngle}\,
{\rm e}^{-in\azAngle}\,
f_G(\rho) \,\sin \xi_1 \, \sin \xi_2 \,( i \cos \xi_1 +
\sin \xi_1 \,\cos \xi_2 ) \, .
\label{eq:SU2SphaleronSquare}
\eea
For $\xi_1=\xi_2=\pi/2$, we have the $W^3/Z$-string configuration with the
radial functions satisfying the Nielsen-Olesen differential equations.  On the
boundary of the square spanned by $\xi_1$ and $\xi_2$, all fields are
zero except for the charged scalar $\phi_+$, which takes the form:
\bea
\begin{array}{c l c c l}
\xi_1=0 : & \phi_+ = i v f_H(\rho) \, , & \hspace{2cm}&
\xi_1=\pi : & \phi_+ = -i v f_H(\rho) \, , \cr\cr
\xi_2=0 : & \phi_+ = i v \,e^{-i\xi_1}\, f_H(\rho) \, ,&  \hspace{2cm}&
\xi_2=\pi : & \phi_+ = i v \,e^{i \xi_1}\, f_H(\rho) \, .
\end{array}
\eea
To summarize, we have a two-parameter family of finite-energy string configurations
that we will refer to as the \emph{sphaleron square}. The center of the
square at ($\xi_1=\xi_2=\pi/2$) corresponds to the $W^3/Z$-string discussed
earlier. From the $W^3/Z$-string we may go to any point on the boundary of the
square (say $\xi_1=0$) and then deform $f_H(\rho)$ smoothly to unity everywhere
so as to finally reach a vacuum configuration. The energy remains finite during
the entire deformation. In fact, the energy density on the sphaleron square is
\be
{\mathcal E}_{\rm cl} = \biggl[ \frac{2}{g^2} \left( \frac{f_G'}{\rho}
\right)^2 + \frac{v^2}{\rho^2} \,f_H^2 \left( 1 - f_G \right)^2  \biggr] \,n^2
\,\sin^2 \xi_1 \,\sin^2 \xi_2  +
 v^2 \left( f_H' \right)^2   + \lambda v^4 \left(1 - f_H^2 \right)^2 \, ,
\label{eq:ecl}
\ee
which is maximal at the $W^3/Z$-string, thereby justifying the
``sphaleron'' designation.
Note that it is possible to extend this construction to the full
electroweak theory; this approach will be used in a forthcoming study
on the Z--string in 3+1 dimensions.

\subsubsection{Charged scalar condensate}

We make one simple expansion of the sphaleron square ansatz by
introducing an additional radial function for the upper component of
the scalar field.  We will observe below that
such an extension is necessary to avoid technical problems when
integrating the second order Dirac equation for fermions
in the string background, cf.~sec.~\ref{sec:DiracEq}. In addition, this
modification allows us to probe the stability with respect to
variations in the upper Higgs component. So our full \emph{ansatz} for
the string background in the bosonic sector reads:
\bea
\phi & = & v \, \left(\begin{array}{c}
\displaystyle f_H(\rho)(-i\cos\xi_1-\sin\xi_1\cos\xi_2) +f_P(\rho) \\[1mm]
\displaystyle f_H(\rho)\,\sin \xi_1 \sin \xi_2  {\rm e}^{i n \azAngle}
\end{array}\right)\,,
\nonumber \\[2mm]
\vec{W}^3&=& \frac{2n}{g \rho}\hat{\azAngle} f_G(\rho)\, \sin^2 \xi_1
\,\sin^2 \xi_2 \, , \nonumber \\[2mm]
\vec{W}^+ &=& \frac{\sqrt{2}n}{g \rho}\hat{\azAngle}
{\rm e}^{-in\azAngle}\,
f_G(\rho) \,\sin \xi_1 \, \sin \xi_2 \,( i \cos \xi_1 +
\sin \xi_1 \,\cos \xi_2 ) \, ,
\label{eq:SphaleronSquare}
\eea
where $f_P(\rho)$ is non-zero at the origin and goes to zero as $\rho \rightarrow
\infty$.  The magnitude of the Higgs field takes a simple form along the
line  $\xi_2 = \pi/2$,
\be
| \phi |^2 = f_H^2 + f_P^2 \,.
\ee
In the following, we shall restrict our variational search 
to $\xi_2 = \pi/2$. The additional Higgs component then contributes
\be
\Delta{\mathcal E}_{\rm cl}=
v^2\left[f_P^{\prime2}+\frac{n^2}{\rho^2}f_P^2 f_G^2 \,\sin^2\xi_1\right]
+v^4f_P^2\left[f_P^2+2f_H^2-2\right]
\label{eq:Deltaecl}
\ee
to the energy density, eq.~(\ref{eq:ecl}).


\section{Fermions on Strings}
\label{sec:FermionsOnStrings}

As we have seen in the last section, electroweak strings are classically
unstable. It is conceivable, however, that their quantum interactions with
leptons and quarks might lead to stabilization: Since the Higgs condensate is
suppressed in the vortex core, the fermions become effectively massless
on the string, so energy can be gained by populating these states. If
the energy gain is large enough, the string ceases to decay and it
becomes a stable multi--quark object with a potentially
rich phenomenology~\cite{Kibble,Hindmarsh:1994re}. 

To treat the fermions consistently, we not only have to consider the
bound states but also the full Dirac sea, {\it i.e.\@} the fermion vacuum
polarization, $E_{\rm vac}$. This quantity is the most cumbersome part
of the computation and to our knowledge, it has not been calculated
properly before. The main difficulty arises because the (string)
background is of non--perturbative nature while the renormalization
conditions are formulated within perturbation theory.
In this chapter we will discuss our approach to this issue in
greater detail.

Let us assume that $N_f$ fermions (with perturbative mass $m_f$) are
trapped along the string by occupying $N_f$ of the bound states that
are induced by the string background. These bound states are characterized
by energy eigenvalues\footnote{Our background profiles lead to $CP$ invariant
interactions for the fermions, cf.~sec.~\ref{sec:DiracEq}. It is therefore
sufficient to consider non--negative energy eigenvalues and positive
fermion numbers $N_f$.} $0 \le \epsilon_i < m_f$.
The explicit occupation of $N_f$ bound states then contributes
\be
 E_{\rm occ}^{(N_f)}=\sum_{i=1}^{N_f}\epsilon_i
\label{eq:Eocc}
\ee
to the total energy. It is, of course, clear that the minimal total energy is
found if we occupy the $N_f$ \emph{lowest} bound states in eq.~(\ref{eq:Eocc}).
The total energy associated with fermion number $N_f$ is then
\be
E_{\rm eff}^{(N_f)} = E_{\rm cl} + E_{\rm occ}^{(N_f)} + E_{\rm vac} \, .
\label{eq:Eeff}
\ee
We emphasize that $E_{\rm occ}^{(N_f)}$ and $E_{\rm vac}$ are formally of
the same order in $\hbar$, and the inclusion of either thus requires the other.
The effective energy $E_{\rm eff}^{(N_f)}$ must be compared with the mass of
$N_f$ elementary (perturbative) fermions, {\it i.e.\@} stabilization occurs if
\be
E_{\rm eff}^{(N_f)} < m_f N_f \, .
\label{eq:stable}
\ee
Our main goal is to perform a variational search in our string \emph{ansatz}
space, so as to minimize $E_{\rm eff}^{(N_f)}$ until eq.~(\ref{eq:stable}) holds.
It is immediately evident that certain situations will be favorable
\emph{a priori}:
\begin{enumerate}
\item The perturbative fermion should be as heavy as possible; we will
therefore concentrate on the top quark contribution to  $E_{\rm eff}^{(N_f)}$.
\item If fermions carry an internal degree of freedom ({\it e.g.\@}
\emph{color}) and $N_C$ is the corresponding degeneracy, then both
$E_{\rm occ}^{(N_f)}$ and $E_{\rm vac}$ acquire a factor $N_C$,
whereas $E_{\rm cl}$ does not. Thus a large color degeneracy
enhances the quantum piece of the total energy.  
The fermion number with internal degeneracy is then $(N_f\,N_C)$,
and the stability criterion eq.~(\ref{eq:stable}) becomes
$E_{\rm eff}^{(N_f)} < m_f\, (N_f\, N_C)$.
\end{enumerate}

\subsection{The Theory}

We ignore inter--generation mixing and set the CKM matrix to unity.
For simplicity, we shall also assume that fermions within an isospin doublet
are degenerate.  This simplifying assumption is violated in reality
(in particular for the heavy quarks), but it gives us an additional
symmetry which we need to make the calculation of $E_{\rm vac}$ manageable.
Let us therefore consider one degenerate isospin doublet $\Psi$ of
heavy fermions ({\it e.g.\@} the top--bottom pair).  In $D=2+1$ the
action for the doublet is
\be
S_F[ \Psi, \Phi, W_\mu] = \int d^3 x \left[ \bar{\Psi} i \gamma^\mu
D_\mu P_L \Psi +\bar{\Psi} i \gamma^\mu \partial_\mu P_R \Psi
- f \bar{\Psi} \left( \Phi P_R + \Phi^\dag P_L \right) \Psi
\right] \, ,
\label{eq:SF}
\ee
where
\be
P_{L,R}=\frac{1\mp\gamma^5}{2} \quad{\rm and}\quad
\Phi=\pmatrix{ \phi_0^{*} & \phi_+ \cr -\phi_+^* & \phi_0}\,.
\label{eq:matrixPhi}
\ee
The degeneracy of the isospin doublet is reflected by the appearance
of a single Yukawa coupling constant, $f$.  The corresponding fermion mass,
\begin{equation}
m_f = f v
\end{equation}
is generated as usual from spontaneous symmetry breaking via the Higgs \vev.
The coupling to the gauge fields appears through the covariant
derivative, which takes the same form as eq.~(\ref{eq:covariantDerivative}),
except that the gauge fields $W_\mu$ only couple to the left-handed fermions
$P_L \Psi$ (as in the full electroweak theory).  Finally, we employ 4-component
Dirac spinors even in $D=2+1$ to mimic the full $D=3+1$ calculation as
closely as possible. Including the two isospin degrees of freedom, our
fermion spinors $\Psi$ thus have 8 complex components.

\subsection{The Effective Energy}

In path integral language, the vacuum polarization energy is obtained
by integrating out the heavy fermion doublet. For this purpose, we decompose
the fermion action (\ref{eq:SF}) into the free part and the
interaction with the bosonic background fields. The latter enters in
the form of a background potential, $U = U(W,\Phi)$,
\be
S_F[ \Psi, W_\mu,\Phi]=
\int d^3x\left[\bar{\Psi}(i\gamma^\mu\partial_\mu-m_f\ID)\Psi
-\bar{\Psi}\,U(W,\Phi)\,\Psi\right]\,,
\label{eq:DefU}
\ee
which may easily be extracted from eqs.~(\ref{eq:SF}) and~(\ref{eq:matrixPhi}).
If we normalize the fermion loop contribution to the boson action by
subtracting off the result in the free non-interacting case, a
formal path integral expression would be
\begin{equation}
S_q[W,\Phi] \equiv S_q[U] = (-i)\, \log \,\frac
{\int d[\Psi,\bar{\Psi}]\,\exp(i S_F[\Psi,U])}
{\int d[\Psi,\bar{\Psi}]\,\exp(i S_F[\Psi,U=0])}\,.
\label{eq:Sq}
\end{equation}
In the language of Feynman diagrams, eq.~(\ref{eq:Sq}) equals the sum of all
graphs with one fermion loop and an arbitrary number of $(-i U)$ insertions,
{\it i.e.\@}
\begin{equation}
S_q[U] = \sum_{n=1}^\infty S_{\rm FD}^{(n)}[U]\,.
\label{eq:Sq1}
\end{equation}
Notice that the $n=0$ term (corresponding to a cosmological constant)
is not included due to our normalization eq.~(\ref{eq:Sq}). As it stands,
eq.~(\ref{eq:Sq1}) is, of course, ill-defined
due to severe UV divergences in the low-order Feynman diagrams. We can identify
these divergent graphs by power counting and assign a definite value to them
using a specific regularization method that introduces a regulator
(for instance a momentum cut--off, $\Lambda$, or the deviation, $\epsilon$,
from the physical dimension). Next, we cancel the divergences by
adding counterterms to the initial (tree-level) boson Lagrangian.
If the regularization method respects the \emph{gauge invariance} of the
fermion loop, the counterterms will be gauge invariant, too.\footnote{This
statement is more subtle when boson loops are included. In this case, the gauge
fixing requirement only leaves a global BRS invariance of the bare action, which
does \emph{not} translate directly into BRS invariant counterterms.}
Moreover, since our model is renormalizable, the counterterms are
local monomials in the boson fields of mass dimension four or
less. With these constraints, the most general counterterm Lagrangian
in the Higgs-gauge sector reads
\be
\mathscr{L}_H^{\rm (ct)} =
c_1\tr\left(G^{\mu\nu}G_{\mu\nu}\right) +
c_2\left[D^{\mu}\phi\right]^{\dag}D_{\mu}\phi+
c_3\left[\left(\phi^{\dag}\phi\right)-v^2 \right] +
c_4\left[\left(\phi^{\dag}\phi\right)-v^2\right]^2 \, .
\label{Lct}
\ee
After adding this term to the initial Lagrangian \eq{eq:HiggsLagrangian} and
replacing all fields and couplings by the renormalized ones,\footnote{If the
context does not lead to confusion, we will use the same notation for
renormalized and bare fields/parameters.} the fermion loop calculation is
now based on the starting bosonic Lagrangian
$\mathscr{L}_H^{\rm (ren)} + \mathscr{L}_H^{\rm ct}$. The counterterm
coefficients $c_i$ are of order ${\mathcal O}(\hbar)$ and enter
the one-loop effective action directly,
\be
S_{\rm eff}^{\rm (ren)}[W,\Phi] = S_H[W,\Phi]
+ \left( S_H^{\rm (ct)}[W,\Phi] +\sum_{n=1}^{\infty}S_{\rm FD}^{(n)}[U] \right)
\equiv S_H[W,\Phi] + S_{\rm vac}[W,\Phi]\,.
\label{eq:SeffFDren}
\ee
We can adjust the coefficients $c_i$ such that they cancel the
divergences in the low order Feynman diagrams. The remaining (finite)
arbitrariness is fixed by imposing specific renormalization conditions
on Green's functions at prescribed external momenta (discussed in more
detail in section \ref{sec:counterterms}). Although there is no unique
specification of the theory by renormalized parameters, we prefer to
choose a scheme where these quantities take \emph{empirical} values
that may, in principle, determined from experiment; a complete list is
given in the next subsection.

The discussion above concentrated on the effective action. For static
boson fields, all these statements translate rather trivially into
expressions for the energy: We simply have to replace
$S \to E \equiv - \lim_{T \to \infty} S/T$, where $T$ is the time
extension of our fields. For instance, the one-loop effective action gives
rise to a one-loop quantum energy
\begin{equation}
E_{\rm eff}[W,\Phi] \equiv E_{\rm cl}[W,\Phi] + E_{\rm vac}[W,\Phi]
= -\lim_{T\rightarrow\infty} T^{-1}\, S_{\rm eff}^{\rm (ren)}[W,\Phi]
\label{Eeff}
\end{equation}
where $E_{\rm cl}$ refers to the classical energy of the \HiggsGauge
sector, {\it cf.\@} \eq{eq:ecl}. The vacuum polarization piece in \eq{Eeff}
has two parts, the total fermion loop correction (\ref{eq:Sq1})
(with the replacement $S\to E=-S/T$) and the counterterm contribution
\be
E_{\rm ct}[W,\Phi]  = \int d^2x \Biggl\{
-c_1\tr\left(W_{i j}W_{i j}\right)
+c_2\left[D_{i}\phi\right]^{\dag} D_{i}\phi
-c_3\left[\left(\phi^{\dag}\phi\right)-v^2\right]
-c_4\left[\left(\phi^{\dag}\phi\right)-v^2\right]^2 \Biggr\}\,.
\label{eq:Ect}
\ee
It should be emphasized again that only the \emph{sum} of these two parts
is finite as the regulator is removed, {\it e.g.\@} $\Lambda\to\infty$ or
$\epsilon\to0$. One of the main advantages of our method is that the two
contributions to $E_{\rm vac}$ are manifestly combined such that no
explicit cutoff is required even at intermediate stages of the calculation.

Besides the classical, vacuum polarization and counterterm piece,
a given configuration with fermion number $N_f$ will also have to
include a contribution from explicitly occupied bound states,
\begin{equation}
E_{\rm eff}^{(N_f)}[W,\Phi] = E_{\rm cl}[W,\Phi] + E_{\rm occ}^{(N_f)}[W,\Phi] +
E_{\rm vac}[W,\Phi]\,.
\label{eq:main}
\end{equation}
This is the quantity to be used for the main stability criterion
eq.~(\ref{eq:stable}). In the following subsections, we will discuss the
individual ingredients into our main formula (\ref{eq:main}) in more
detail.

To close this section, a brief comment is in order: We are well aware that a
consistent one-loop calculation must include gauge and Higgs field loops, too.
However, we choose to ignore these contributions in the present
calculation, since our main motivation is slightly different:  We are
mainly concerned with the study of fermion effects on bosonic
backgrounds.  A consistent treatment of the fermion sector requires
the inclusion of both bound state and sea quark contributions.
Although bosonic loops also appear at the same order in $\hbar$, they
are suppressed compared to the fermionic fluctuations by a factor of
the number $N_C$ of internal fermion degrees of freedom ({\it e.g.\@}
color).  Our approximation is relevant when $N_C$ can be assumed to be
large, becoming exact in the limit $N_C \to\infty$, since all boson
lines in Feynman diagrams are suppressed in this case.

\subsection{Theory Parameters}

As mentioned earlier, we choose to define our model by renormalized
parameters that have, in principle, a direct relation to electroweak
phenomenology. However, our model is defined in two space dimensions so that
the (dimensionful) gauge coupling constant $g$ can no longer be expressed
in terms of the Fermi constant $G_F\approx 10^{-5}m_{\rm proton}^{-2}$
via the standard relation $G_F/\sqrt{2}=g^2/M_{\rm W}^2=2/v^2$. To see
which value of $v$ should be chosen in $D=2+1$, consider the classical
energy (or energy per unit length) in eq.~(\ref{eq:ecl}). In any number of
dimension this quantity takes the form $v^2$ times a dimensionless
function involving only ratios of masses. Thus the mass dimension comes
from the prefactor $v^2$ only, which must therefore be adjusted such that the
known $D=3+1$ value of $v^2$ is multiplied by an appropriate length scale.
Since we relate all dimensionful quantities to the mass $m_f$ of the heavy
fermion, one might expect that the appropriate length scale is the Compton
wave length of the heavy fermion, $2\pi/m_f$. This is almost correct; the
actual calculations of quantum energies for the QED flux tube
(eqs.~(42) and~(43) in ref.~\cite{Graham:2004jb}) show that the proper
scaling factor is in fact \emph{half} the Compton wave--length. Thus,
the physically sensible choice for $v^2$ is
\be
v_{D=2+1}^2=\frac{\pi}{m_f}v_{D=3+1}^2=\frac{\pi}{2\sqrt{2}G_Fm_f}\,.
\label{eq:vev3d}
\ee
For the remaining parameters, {\it i.e.\@} the masses that determine
the coupling constants $f$, $g$ and $\lambda$, we can directly take
the values suggested by electroweak phenomenology:
\be
m_f = fv_{D=3+1}=170\,{\rm GeV}\,,\quad \frac{gv_{D=3+1}}{\sqrt2}=80\,{\rm GeV}
\quad {\rm and}\quad 2v_{D=3+1}\sqrt{\lambda}=115\,{\rm GeV}\,.
\label{eq:massparameters}
\ee
We have identified the heavy fermion doublet with the top/bottom pair
and set $m_f$ to the top quark mass. In our numerical studies, we have also
experimented with a very heavy fermion to investigate possible decoupling
scenarios.

\subsection{The Counterterms}

\label{sec:counterterms}

The counterterm coefficients $c_i$ in (\ref{eq:Ect}) must be chosen to
cancel the divergences in low-order Feynman diagrams. In $D=2+1$, only the
first two diagrams (with $n=1$ and $n=2$ insertions of the background potential
$-iU$) are divergent and it is readily seen that the terms in (\ref{eq:Ect})
are sufficient to render the sum of the $n=1,2$ diagrams --- and thus the
entire vacuum polarization part $E_{\rm vac}$ --- finite. The remaining
arbitrariness in the definition of the $c_i$ is fixed by conventional
renormalization conditions (see, for instance, ref.~\cite{KhemaniDecoupling}):
\begin{itemize}
\item[a.]
We start by choosing the so--called \emph{no-tadpole} condition.
This ensures that neither the VEV $\langle\phi\rangle=v\,(0\,,\,\,1)^T$
nor the perturbative fermion mass $m_f$ receive radiative corrections.
This condition determines $c_3$.
\item[b.]
We fix the pole of the Higgs propagator to be at its tree level mass,
$M_H$, with residue one. These conditions yield $c_2$ and $c_4$.
\item[c.]
There are various choices to fix the remaining coefficient $c_1$. We choose
to set the \emph{residue} of the pole of the gauge field propagator to one
in unitary gauge. The \emph{position} of that pole, {\it i.e.\@} the mass
of the gauge field, $M_{\rm W}$, is then a prediction.
\end{itemize}
The resulting counterterm coefficients $c_i$ are listed in the Appendix.
As explained under item c., the mass of the gauge fields
is constrained by the other model parameters when fermion
loops are included. With our choice of renormalization conditions,
$M_W$ is the solution to the implicit equation
\be
\xi_{\rm W}^2=\frac{g^2v^2}{2m_f^2}
(1+\frac{f^2}{\pi m_f}I_2(\xi_{\rm W}))
+\frac{g^2}{16\pi m_f}(\xi_{\rm W}^2I_1(\xi_{\rm W})+2\bar{I}_1(\xi_{\rm W})),
\label{ModelParamsConstraint}
\ee
where $\xi_{\rm W} = M_{\rm W}/m_f$; the Feynman parameter integrals $I_1$,
$\bar{I}_1$ and $I_2$ are also given in the Appendix.
If we take for $g$ and $m_f=fv$ the values extracted from eqs.~(\ref{eq:vev3d})
and~(\ref{eq:massparameters}), we can solve \eq{ModelParamsConstraint}
numerically to obtain the fermion loop correction to the gauge boson mass.
Due to the condition c. above, this mass corresponds to one-$W$-boson
states in unitary gauge.

Of course, other renormalization conditions are possible. They correspond to
finite changes in the parameters $c_i$ which are exactly calculable and
affect the counterterm contribution \eq{eq:Ect} only. The renormalization group
ensures that such re-parameterizations do not change the physical content of
the model.

\subsection{The Vacuum Polarization Energy}
\label{sec:vpe}

From eq.~(\ref{eq:Sq1}) the one-fermion-loop correction to the bosonic
background energy is an infinite series of Feynman diagrams, which
is impossible to sum in practice. Perturbative methods based on the
lowest order diagrams are ineffective, since the
coupling of the heavy fermion to the boson background is not naturally
small. Instead, we follow the spectral approach introduced in refs.
\cite{PhaseshiftsGeneral,PhaseshiftsFermions} for an \emph{exact}
calculation of the renormalized fermion vacuum polarization.
For a detailed review of this method and a list of further references,
see ref.~\cite{Leipzig}.

The spectral approach is based on the observation that the vacuum polarization
can alternatively be characterized as a sum over the change in the zero-point
energies of the fermion modes due to the background fields. The fermion modes
comprise both the bound states of energy $\epsilon_j$ (with degeneracy $D_j$),
and the scattering modes characterized by a momentum density
of states.  If $\Delta\rho(k)$ denotes the \emph{change} in the density
of the scattering states caused by the non-trivial background,
we have the expression
\be
E_{\rm vac} = -\frac{1}{2}\sum_j D_j(|\epsilon_j|-m_f) -
\frac{1}{2}\int_0^\infty dk \,(\sqrt{k^2+m_f^2}-m_f)\, \Delta\rho(k) +
E_{\rm ct}
\label{Evac2}
\ee
where the degeneracy factor $D_j$ in $D=2+1$ dimensions is unity if the sum
over $j$ runs from $-\infty$ to $\infty$.  The cancellation of the
divergences in the momentum integral by conventional counterterms
$E_{\rm ct}$ is still formal at this stage.

What makes the spectral method effective is the fundamental relation
between the change in the continuum density of states and
the scattering phase shifts $\delta_\ell(k)$ of the Dirac
wave-functions \cite{Schwinger}
\begin{equation}
\Delta\rho(k) = \frac{1}{\pi}\frac{d}{dk}
\sum_\ell D_\ell\,\delta_{\ell}(k) \,.
\end{equation}
The shorthand notation $\ell$ will be used in the following to denote
the collection of all quantum numbers (including the sign of the
energy and generalized angular momentum) that characterize an
individual scattering channel.

Since the Feynman series is an expansion in powers of the interaction
potential \eq{eq:DefU}, the same expansion in the spectral method
suggests a one-to-one correspondence between Feynman diagrams and the
\emph{Born expansion} of the phase shifts.  This reasoning is
essentially correct, though there is a slight subtlety associated with
the string background.  Ignoring this issue for a moment, we may
subtract the lowest orders of the Born series from the phase shifts,
and add back in \emph{exactly} the same quantity in the form of
Feynman diagrams:
\be
E_{\rm vac} = -\frac{1}{2}\sum_j D_j(|\epsilon_j|-m_f)
-\frac{1}{2}\int_0^\infty \frac{dk}{\pi} (\sqrt{k^2+m_f^2}-m_f)
\frac{d}{dk}\,\left[{\delta}(k)\right]_N
+\left[\sum_{i=1}^N E_{\rm FD}^{(i)} + E_{\rm ct}\right] \, ,
\label{eq:Evacgeneral}
\ee
Here, $\left[{\delta}(k)\right]_N$ is a shorthand notation for the sum over
all channels $\ell$ of the full phase shift (including the degeneracy)
minus the first $N$ terms of the respective Born series.
Ref.~\cite{Graham:2002xq} contains a careful derivation of \eq{eq:Evacgeneral}
that starts directly from the definition of the energy-momentum tensor
in field theory.

All three contributions in \eq{eq:Evacgeneral} are separately finite as the
regulator is removed: The bound state is a finite sum, the momentum integral
converges due to the Born subtraction, and the last piece is a conventional
perturbative calculation of the $N$ lowest order Feynman diagrams
(with counterterms to implement the renormalization conditions
discussed above).  For the present model in $D=2+1$, only the first
and second order Feynman diagrams require renormalization, {\it
i.e.\@} $N=2$ is the minimal number of Born subtractions necessary for
a finite calculation. The final expression for the vacuum polarization
energy is thus
\begin{eqnarray}
E_{\rm vac} & = & -\frac{1}{2}\sum_j D_j(|\epsilon_j|-m_f)
-\frac{1}{2}\int_0^\infty \frac{dk}{\pi}\,(\sqrt{k^2+m_f^2}-m_f)
\frac{d}{dk}\,\left[{\delta}(k)\right]_2+E^{(1,2)}+E_{\rm ct}^{(3,4)} \, .
\label{eq:Evac3}
\end{eqnarray}
The first and second order diagrams, combined with the counterterms of the
same order give the finite contribution $E^{(1,2)}$. The third and fourth
order diagrams are \emph{not} divergent in $D=2+1$ (and therefore left as
implicit part of the momentum integral), but we still need finite counterterms 
$E_{\rm ct}^{(3,4)}$ for them to enforce the correct renormalization
conditions. The explicit expressions for $E^{(1,2)}$ and $E_{\rm ct}^{(3,4)}$
can be found in the Appendix.

Let us finally discuss the subtlety associated with the Born approximation in
the string background. It is due to the fact that Feynman diagram
calculations require \emph{all} field derivatives to vanish sufficiently fast
as $\rho\to\infty$. This condition is not met by the vortex, which has winding
and thus nonzero azimuthal derivatives at $\rho\to\infty$.   In $D=2+1$,
only the counterterm coefficient $c_3$ is UV divergent while the remaining
(finite) $c_i$ serve to enforce our renormalization scheme. The counterterm
multiplied by $c_3$ involves only $(\phi^{\dagger}\phi)$, so instead
of doing the full Born subtraction and the corresponding diagrams, we
can perform this piece of the calculation in a \emph{fictitious}
background with the same value $(\phi^{\dagger}\phi)$ as our initial
vortex background. On the sphaleron square, we can, for example, choose
the corresponding boundary point with $\xi_1=0$ and $\xi_2=\pi/2$, {\it cf.\@}
Sec.~\ref{sec:interpolation}, where all gauge fields vanish and
$\phi$ has no azimuthal dependence at $\rho\to\infty$. For this
fictitious background, the subtraction/diagram part of the calculation
is straightforward.

\subsection{The Dirac Equation}
\label{sec:DiracEq}
\noindent
The final ingredients in our computational method are the 
equations of motion for the fermion fields and the associated
techniques used to extract the phase shifts in the boson background. 

\subsubsection{First-order equations}

We start from the Dirac equation in $2+1$ dimensions, which we write in the
form\footnote{With four-component Dirac spinors, the same equation also holds
in $D=3+1$, provided that all fields are translationally invariant
along the string (which we put along the $z$--axis) and the gauge
fields have vanishing $z$-components. This means that the bound states
and phase shifts computed in this section are also appropriate for
$D=3+1$ and the trivial space dimension can be managed by the
interface formalism~\cite{GrahamInterface}. Complications arise in
$D=3+1$ only because the renormalization procedure is more elaborate.}
\be
h_D \Psi_{\omega, N_\ell} (\rho, \azAngle) = \omega \Psi_{\omega, N_\ell}
(\rho, \azAngle) \, ,
\label{eq:DiracEq}
\ee
The single particle Dirac operator can be read off from \eq{eq:SF},
\be
h_D = -i\alpha^i\partial_i + \gamma^0 f \left( P_R \Phi + P_L \Phi^\dag
 \right) - g\,\alpha^i \biggl[ Z_i \,Q_Z 
+ P_L (W_i^+T^- + W_i^-T^+) \biggr] \, ,
\ee
where $\alpha^i=\gamma^0\gamma^i$ and $i=1,2$. Our vortex configurations are
special cases of a class of background fields which show an
angular dependence similar to the Nielson--Oleson vortex
of winding number $n$, 
\be
\phi_0 \propto  {\rm e}^{i n \azAngle} \, , \qquad
\phi_+  \propto  1 \, , \qquad
\vec{W}^3  \propto  \hat{\azAngle} \, , \qquad
\vec{W}^+=\left(\vec{W}^-\right)^*
\propto  {\rm e}^{-i n \azAngle} \hat{\azAngle} \, ,
\label{eq:symmetry2}
\ee
This class of configurations exhibits a symmetry: Since the heavy fermion
doublet is degenerate, the operator
\be
K = L_3 + S_3 - n T_3 \gamma_5
\label{eq:symmetry1}
\ee
commutes with the Hamiltonian $h_D$. Its eigenvalues
$N_\ell = \ell+\half+\frac{n}{2}$ may therefore be used to label the
fermion modes and channels in a vortex background of the specified form.

As explained earlier, we want to keep our computation as close as possible to
the $D=3+1$ case, and therefore employ Dirac 4-spinors even in $D=2+1$.
This redundancy implies the existence of a matrix $\alpha^3$ with the
property $\{\alpha^3, h_D \}=0$. For every eigenspinor $\Psi$ of the Dirac 
operator
with eigenvalue $\omega$, there exists an corresponding state $\alpha^3 \Psi$
with eigenvalue $-\omega$, and the spectrum of $h_D$ is manifestly symmetric
about zero.  

For the actual calculation we choose the chiral representation of the
Dirac matrices (with $\gamma^5= {\rm diag}(-1,-1,1,1)$ and 
$S_3 = {\rm diag}(1,-1,1,-1)/2$) 
and parameterize the eigenspinors by eight functions $y_1,\ldots,y_8$ that only 
depend
on the radial coordinate $\rho$,
\begin{equation}
\Psi_{\omega, N_\ell}(\rho, \azAngle) = \left( \begin{array}{r@{\,\,+\,\,}l}
y_5 \,e^{i \ell \azAngle}\,|\half\rangle &
y_7 \, e^{i (\ell + n)\azAngle}\,|-\half\rangle \\[2mm]
iy_6 \,e^{i (\ell + 1) \azAngle}\,|\half\rangle &
iy_8 \,e^{i(\ell+n+1)\azAngle}\,|-\half\rangle \\[2mm]
y_1 \,e^{i (\ell + n) \azAngle}\,|\half\rangle &
y_3 \,e^{i \ell \azAngle}\,|-\half\rangle \\[2mm]
iy_2 \,e^{i (\ell + n +1) \azAngle}\,|\half\rangle &
iy_4 \,e^{i ( \ell + 1) \azAngle}\,|-\half\rangle
\end{array}  \right)
\label{eq:Psi_param}
\end{equation}
The kets in this expression are eigenvectors of $T^3 = \tau^3/2$.
We may re-cast the $D=2+1$ Dirac Hamiltonian in $2\times2$ block form as
\be
h_D=\pmatrix{h_{11} & h_{12} \cr h_{21} & h_{22} } \,.
\label{eq:Dirac2x2}
\ee
In the chiral representation of the Dirac matrices (and in temporal gauge)
these $2\times2$ blocks have a very simple dependence on the background fields,
\bea
h_{11}&=&i\sigma_i\partial_i+g\,\sigma_i\left\{Z_i\,T^3
+ W_i^+T^- + W_i^-T^+\right\}\,, \nonumber \\
h_{22}&=&-i\sigma_i\partial_i\,, \nonumber \\
h_{12}&=&f\Phi\,, \\
h_{21}&=&f\Phi^\dag \, . \nonumber
\eea

\subsubsection{Second-order Equations}

To extract the scattering data, the second order form of the equations
of motion is more appropriate. We find these by substituting the 
Dirac operator \eq{eq:Dirac2x2} into the first order form
\eq{eq:DiracEq} which eliminates half of the field components (while
doubling the order of the differential equation).  We choose to
eliminate the left--handed fields, which are the upper components in
the chiral representation. The resulting equation of motion
for the right-handed (lower) Weyl spinor $\Psi_R$ reads
\be
h_{21} \left[ (h_{11}-\omega) h_{21}^{-1} (h_{22} - \omega) - h_{12} \right]
\Psi_R = 0 \, .
\label{eq:secDirac}
\ee
These are two complex second--order equations for the two complex 
components of $\Psi_R$. Upon inserting the parameterization 
eq.~(\ref{eq:Psi_param}), we obtain four second--order equations 
for the radial functions appearing in $\Psi_R$,
\begin{equation}
\sum_{j=1}^{4} \left\{ \mathscr{D} (\rho) + \mathscr{N}(\rho)\,\frac{d}{d \rho} +
\mathscr{M} (\rho) \right\}_{ij}y_j(\rho) = 0 \, .
\label{eq:DiracDNM}
\end{equation}
with $k^2=\omega^2-m_f^2$.
The kinetic operator $\DD$ contains a radial piece from the 2D Laplacian,
as well as a generalized centrifugal barrier $\OO$,
\be
\mathscr{D}(\rho)=\mathbbm{1}\left(\frac{d^2}{d\rho^2} +
\frac{1}{\rho}\frac{d}{d\rho}+k^2\right)-\frac{1}{\rho^2}\,\mathscr{O}
\, , \qquad
\mathscr{O}={\rm diag}\left[(\ell + n)^2\,,(\ell+n+1)^2\,,\ell^2\,,
(\ell+1)^2\right]\,.
\ee
The matrices $\mathscr{N}(\rho)$ and $\mathscr{M}(\rho)$
contain the bosonic background potential. The explicit form of the matrix
elements is rather lengthy, particularly for the non-trivial background
\emph{ans\"atze} in our numerical treatment; we refrain from presenting them
in any detail. At large radii $\rho\to\infty$, the background
potential vanishes and $\NN(\rho)\to0$ as well as $\MM(\rho)\to0$, and
the four differential equations decouple.

The bound state energies $\epsilon_j$ can be computed directly from
eq.~(\ref{eq:DiracEq}) by shooting for normalizable solutions that are regular
at the origin and decay exponentially at large radii $\rho\to\infty$.
By contrast, the phase shifts, which are extracted from
eq.~(\ref{eq:DiracDNM}), require a few more manipulations to put the
calculation in a manageable form.

We have a four-channel scattering problem that can be treated in the usual
fashion: The full eigenspace to a given energy $\omega$ is decomposed
into channels labeled by the integral angular momentum $\ell$. In each
such channel, we have four independent solutions
$\vek{y}(\rho)$ to eq.~(\ref{eq:DiracDNM}), which we may combine in the
\emph{rows} of a matrix $\YY$. Thus, $\YY_{ij}(\rho,k)$ is the radial function
$y_j(\rho)$ in the $i^{\rm th}$ linearly independent solution at a fixed
momentum $k$ and channel number $\ell$ (we suppress $\ell$ in the following).

Let us analyze the free case $\MM = \NN = 0$ first. The solutions with
outgoing spherical wave boundary conditions at $\rho\to\infty$, are
$\YY(\rho,k) = \HH(k\rho)$ with
\be
\HH(x)={\rm diag}\left[h_{|\ell+n|}^{(1)}(x) \, ,\,
h_{|\ell+n+1|}^{(1)}(x)\, , \,h_{|\ell|}^{(1)}(x) \, ,\,
h_{|\ell+1|}^{(1)}(x)\right]
\ee
given explicitly in terms of Hankel functions. Similarly, for $\MM ,
\NN \ne 0$, the scattering states
$\vek{y}(\rho)$ are solutions of the fully interacting equations of
motion (\ref{eq:DiracDNM}), which tend to specific free outgoing waves
of the same energy at $\rho\to\infty$. Combining them in a $(4 \times
4)$ matrix as before, we may split off the asymptotic form by the
ansatz $\YY(\rho,k) = \FF(k,\rho)\cdot \HH(k\rho)$.  Inserting into
eq.~(\ref{eq:DiracDNM}), we find the differential equation for the
Jost-like matrix $\FF(k,\rho)$,
\be
\FF^{\prime\prime}+\frac{1}{\rho}\FF^\prime+2 \FF^\prime \LL^\prime+
\frac{1}{\rho^2}[\FF,\OO]+ \NN(\FF^\prime+\FF \LL^\prime)+ \MM \FF =0 \, ,
\ee
where $\LL(k\rho)\equiv \ln \HH(k\rho)$ and primes denote
derivatives with respect to the radial coordinate.
If we impose the boundary conditions
\[
\FF(k,\rho) \to \mathbbm{1},\qquad\quad\FF'(k,\rho) \to 0
\]
at $\rho\to\infty$, the corresponding solution matrix
$\YY(\rho,k) = \FF(k,\rho)\cdot \HH(k\rho)$ clearly describes
outgoing spherical waves. What is unusual about the string background is
that the matrices $\MM$ and $\NN$ may actually be complex. As a consequence,
the conjugated matrix $\YY^\ast$ does \emph{not} describe \emph{incoming}
spherical waves; in fact, it is not even a solution of the equations
of motion.
Instead, the scattering states corresponding to incoming waves, which we
denote by $\overline{\YY}$ in the following, must be computed separately by
an appropriate change in the Hankel functions $\HH \to \overline{\HH}$.
The standard scattering analysis now proceeds as usual with $\YY^\ast$
replaced by $\overline{\YY}$.\footnote{Alternatively, we could rewrite the
complex $(4\times 4)$ problem as a real $(8 \times 8)$ problem. The phase
shifts -- or more precisely the sum of eigenphaseshifts in each
channel -- extracted in this way are exactly \emph{twice} as
large as the ones from the initial $(4\times 4)$ problem. This
statement is obvious in the case that the matrices $\MM$ and $\NN$ are 
real, and we have checked it numerically in other cases.}

The scattering wavefunction can now be written as
\begin{equation}
\YY_{\rm sc}(\rho)=-\overline{\YY}(\rho) + \YY(\rho) \cdot \Sc(k)\, .
\end{equation}
Requiring that $\YY_{\rm sc}(\rho)$ be regular at the origin yields
a relation for the scattering matrix,
\begin{equation}
\Sc(k)=\lim_{\rho\to0}\YY^{-1}(\rho)\cdot\overline{\YY}(\rho)=
\lim_{\rho\to0} \HH^{-1}(k\rho)\cdot \FF^{-1}(k,\rho)\cdot \overline{\FF}(k,\rho)
\cdot \overline{\HH}(k\rho)\,.
\end{equation}
For the spectral method (\ref{eq:Evac3}), we need the sum of the eigenphase
shifts obtained from $\Sc$,
\begin{equation}
\delta(k) = \frac{1}{2i}\Tr \ln \Sc(k) =
\frac{1}{2i}\lim_{\rho \to 0}\Tr\ln \left( \FF^{-1}(k,\rho)\cdot
\overline{\FF}(k,\rho)\right) \, .
\label{eq:deltaFF}
\end{equation}
It should be noted that the fermion flux in the chiral basis is not
orthonormal and the matrix $\Sc$ is hence not unitary. However, the
physical scattering matrix can be obtained from $\Sc$ by a change of basis
and proper normalization. Both manipulations do not alter the sum of 
eigenphase shifts, which is the relevant quantity here.

In our numerical calculations, we do not use eq.~(\ref{eq:deltaFF})
directly; instead we set
\begin{equation}
\delta(k,\rho) \equiv \frac{1}{2i}\Tr\ln \left( \FF^{-1}(k,\rho)
\cdot\overline{\FF}(k,\rho) \right) \, ,
\end{equation}
such that $\delta(k)=\delta(k,0)$. To compute this function, we
integrate the differential equation
\begin{equation}
\frac{\partial \delta (k,\rho)}{\partial \rho} =
- \frac{1}{2i}\,\Tr\,\left[ \FF'\cdot \FF^{-1}
- \overline{\FF}'\cdot\overline{\FF}^{-1}\right]\,,\qquad\qquad
\lim_{\rho\to\infty} \delta(k,\rho) = 0
\end{equation}
along with the system for $\FF$ and $\overline{\FF}$ from $\rho=\infty$ to $0$.
This procedure results in a smooth function of both $\rho$ and $k$,
which avoids ambiguous jumps of (multiples of) $\pi$ and allows to
take the limit $\rho\to 0$ easily. 

Notice finally that channels which involve Hankel functions of order
$\ell=0$ are afflicted by a further numerical subtlety.
Eq.~(\ref{eq:deltaFF}) essentially requires to separate the regular and
irregular pieces as $\rho\to0$. For $\ell=0$ this amounts to
distinguishing a logarithmic behavior from a constant,
which is difficult to handle numerically. In this case, we switch to a
radial function analogous to  $\delta(k,\rho)$, which is defined in terms
of the radial \emph{derivatives} of $\FF(k,\rho)$ and $\LL(k,\rho)$.
We have checked the numerical stability of this treatment thoroughly.

\subsection{Derivative expansions}

\noindent
To check our numerical results, it is helpful to have analytic
approximations to the vacuum polarization energy. The lowest orders
of a \emph{perturbative} expansion are already contained in our main
formula \eq{eq:Evac3}; to check the phase shift piece of the calculation,
we would therefore have to compute (many) higher order diagrams, which is
very cumbersome. A better strategy is to employ a \emph{derivative expansion}
which is expected to work well for reasonably broad background profiles.

The leading order of the gradient expansion is the effective potential,
$V_{\rm eff}$, which contains no derivatives of the boson background
fields. Together with gauge invariance, this restriction rules out any dependence
on the gauge fields, so $V_{\rm eff}$ is a functional of the Higgs
field alone.  In $D=2+1$ dimensions, $V_{\rm eff}$ is formally given by
\be
-\int d^3 x \,V_{\rm eff}[\phi] = -i\,{\rm Tr}\, {\rm ln}
\left(i\partial \hskip -0.55em / -f \Phi_5\right)
\label{eq:veff1}
\ee
for a \emph{constant} background $\Phi_5=\Phi P_L+\Phi^\dagger P_R$,
cf.~\eq{eq:matrixPhi}. The functional trace is divergent and
needs to be renormalized by adding the same counterterms as were
determined in section \ref{sec:counterterms}. To get definite expressions,
we employ dimensional regularization and write
\be
-\int d^3 x \,\frac{\partial V_{\rm eff}[\phi]}{\partial f}
=(-i) f\,{\rm Tr}\,\left\{\left[\partial^2+f\Phi_5\Phi_5^\dagger\right]^{-1}
\Phi_5\Phi_5^\dagger\right\} \to
i\,f\, \int d^D x\int \frac{d^Dk}{(2\pi)^D}\,{\rm tr}\,
\left\{\left[k^2-f\Phi_5\Phi_5^\dagger\right]^{-1}
\Phi_5\Phi_5^\dagger\right\}\,.
\label{eq:veff2}
\ee
The analytic continuation of the integral on the rhs to $D=3$ is finite,
so that one would expect to need no renormalization at all. 
However, the integral is manifestly convergent only in
$D=1$ and the continuation to $D=3$ is unambiguous only if the divergence
in the even dimension $D=2$ is canceled by counterterms. 
Only the sum of the integral and counterterm is analytic in the range
$D=1$ to $D=3$ and we \emph{must} renormalize for all $D \ge 2$.
Furthermore, we need to impose renormalization conditions that match our
previous scheme if we want to compare the results of the derivative expansion
with the spectral method. Upon integrating eq.~(\ref{eq:veff2})
with respect to the Yukawa coupling and fixing the integration
constant via the normalization $V_{\rm eff}[v] = 0$, we obtain the result
\be
V_{\rm eff}[\phi]=\frac{2}{3\pi}f^3\,
\left[\left(\phi^\dagger\phi\right)^{3/2}-v^3\right]
-\frac{vf^3}{\pi}\,\left[\phi^\dagger\phi-v^2\right]
-c_4\left[\phi^\dagger\phi-v^2\right]^2\,.
\label{eq:veff3}
\ee
The explicit expression for the counterterm coefficient $c_4$ is
rather lengthy; it can be read off from \eq{eq:countcoef} in
the appendix. Notice that the effective potential is
minimized for configurations with $\phi^\dagger\phi=v^2$;
this is, of course, a direct consequence of our \emph{no-tadpole}
renormalization condition.

For the next-to-leading order in the derivative expansion,
there are well-known techniques \cite{Aitchison:1984ys,Aitchison:1985pp}
that apply straightforwardly in the case of a pure Higgs
background. On the sphaleron square, these are the configurations with
$\xi_1 = 0$. We will see in the next section that such backgrounds
play a prominent role. The result for the effective Lagrangian is
\be
\LL_{\rm eff}^{(2)}=\frac{f}{16\pi}
\left\{\left(\phi^\dagger\phi\right)^{-1/2}
\left(\partial_\mu\phi\right)^\dagger
\left(\partial^\mu\phi\right)
-\frac{1}{3}\left(\phi^\dagger\phi\right)^{-3/2}
\partial_\mu\left(\phi^\dagger\phi\right)
\partial^\mu\left(\phi^\dagger\phi\right)\right\}\,,
\label{eq:leff1}
\ee
where the superscript indicates that this term contains two derivatives
of the Higgs background. Again, the expression \eq{eq:leff1} is
ultra-violet finite in $D=2+1$, but we need to add the two
derivative counterterms, \eq{Lct}, to match our previous renormalization
conditions.

In a gauge invariant expansion, the leading order effect of the gauge
bosons is to turn the ordinary derivatives in eq.~(\ref{eq:leff1}) into
\emph{covariant derivatives} of the form
\eq{eq:covariantDerivative}.\footnote{There are various approaches
to counting derivatives in a covariant expansion; some of these schemes
expand in the gradients of the magnetic field, so the leading term will
contain $\vek{B}$ (and hence the derivatives of the gauge field) at all orders.
For our purposes, the simple expression (\ref{eq:Evacapprox}) is sufficient.}
Inserting the explicit form \eq{eq:SU2SphaleronSquare} of the static
vortex background, our final expression for the derivative expansion
of the vacuum polarization energy is
\begin{eqnarray}
E_{\rm vac}&\approx&2m_f\pi \int_0^\infty xdx\,
\left\{\frac{2}{3\pi}\left[\left(f_H^2+f_P^2\right)^{3/2}-1\right]
+\frac{1}{\pi}\left(1-f_H^2-f_P^2\right)
-\frac{v^4c_4}{m_f^3}\left(1-f_H^2-f_P^2\right)^2\right.\cr
&&\hspace{2.3cm}\left.
+\left[\frac{1}{16\pi}\left(f_H^2+f_P^2\right)^{-1/2}
+\frac{m_f c_2}{f^2}\right]\left[f_H^{\prime2}+f_P^{\prime2}
+\frac{n^2}{x^2}\left(f_H^2(1-f_G^2)+f_P^2f_G^2\right)
{\rm sin}^2\xi_1\right]\right.\cr
&&\hspace{2.3cm}\left.
+\frac{1}{12\pi}\left(f_H^2+f_P^2\right)^{-3/2}
\left(f_Hf_H^\prime+f_Hf_H^\prime\right)^2\right\}\,,
\label{eq:Evacapprox}
\end{eqnarray}
where primes denote derivatives with respect to the dimensionless
variable $x=m_f\rho$.


\section{Numerical Studies}

Our goal is to find a spatially varying configuration of the boson
fields that is energetically favored over the trivial vacuum
configuration with the same fermion number. To this end we scan the
energy surface with respect to variational parameters in specific
\textit{ans\"atze} for the background profiles. The comparison
requires, of course, that both the trivial and non-trivial configuration
have identical fermion numbers. As discussed at the beginning of
section~\ref{sec:FermionsOnStrings},
the vortex background acquires its fermion number by occupying the $N_f$
lowest fermion bound states; its energy must then be compared to the trivial
configuration with energy $N_f m_f$, cf.~eq.~(\ref{eq:stable}). These
considerations become slightly more complicated if internal degrees of freedom
({\it e.g.\@} color) are included, as is discussed in more detail below.

Our variational ansatz is discussed in detail in the next subsection.
Altogether, it involves five parameters; even though this number is
already numerically expensive to be scanned, it is definitely not
exhaustive, so stable configurations can still exist that we have
not found. However, if we find an energetically favored object,
expanding the ansatz will only improve on this result.
We will begin by exploring the parameter dependence
of the various contributions to the total energy and then turn to
the search for bound objects with large fermion numbers.

\subsection{Explicit profiles}

Unless stated otherwise, we shall employ the following
\textit{ans\"atze} for the radial profiles in the string background:
\be
f_H = 1 - e^{-\rho/w_H} \, , \qquad
f_G = 1 - e^{-\left( \rho/w_G \right)^2} \qquad {\rm and}\qquad
f_P = a_{P} e^{-\rho/w_P}.
\label{eq:profiles}
\ee
The constants $w_H$, $w_G$ and $a_P$, together with $\xi_1$,
eq.~(\ref{eq:SphaleronSquare}), are variational parameters
that characterize the background fields, which we will adjust to minimize
the energy functional $E_{\rm eff}^{(N_f)}$, \eq{eq:Eeff}.  The
profile functions $f_H$ and  $f_G$ defined above are constructed such
that the classical energy is  always finite, regardless of the
values chosen for the variational  parameters.  Near the origin, these
radial functions behave as $f_H\propto\rho$ and $f_G\propto\rho^2$
while for large $\rho$ both profile functions tend to
unity. We certainly could introduce additional degrees of freedom 
subject to the boundary conditions.
Introducing additional parameters for the amplitudes of the
exponential functions in $f_H$ and $f_G$ violates this requirement
of finite classical energy, however, and hence we refrain from doing so.

\subsection{Classical Energy}

We first discuss the results for the classical energy $E_{\rm cl}$. We
consider the $W$--string on the sphaleron square,
eq.~(\ref{eq:SphaleronSquare}), with the classical energy
\begin{equation}
E_{\rm cl}=2\pi\int_0^\infty\rho d\rho\,{\mathcal E}_{\rm cl}\,.
\label{eq:eclsphsq}
\end{equation}
The classical energy density, ${\mathcal E}_{\rm cl}$, is given by
eq.~(\ref{eq:ecl}) plus the contribution of the additional scalar field 
$f_P$ as given in eq.~(\ref{eq:Deltaecl}). We always assume unit
winding $n=1$.

We first discuss the case for physical (standard model) parameters
augmented by the choice of the VEV in $D=2+1$ dimensions,
eq.~(\ref{eq:vev3d}).  In what follows all data are measured in units 
of $m_f$ or its inverse.  For the classical energy, we observe a strong 
dependence on the angle $\xi_1$, which parameterizes the interpolation
between the $W$--string and a purely scalar configuration with the same 
Higgs profile $|\phi|^2$.

Even though the mixing angle $\xi_1$ is one of our variational parameters, 
it is fruitful to discuss the energy functional for various \emph{fixed} 
values of $\xi_1$ and attempt to minimize $E_{\rm cl}$ in the space of the 
remaining variational parameters. Note, however, that minimization of 
$E_{\rm cl}$ is not the main goal of our investigations. For small $\xi_1$ 
the gauge field contributions are suppressed and the minimal value for 
$E_{\rm cl}$ is obtained for small Higgs field extensions. In the extreme 
case of no gauge fields ($\xi_1=0$) we expect $E_{\rm cl}\approx6$ as 
$w_H\to0$ and $a_P=0$ from the scaling properties of the 
derivative terms in $\int \rho d\rho\, \mathcal{E}_{\rm cl}$,
{\it cf.\@} eq.~(\ref{eq:ecl}). For
$\xi_1=0.05\pi$, $w_H=1.1$, $w_P=2$ and $a_P=0.1$  we find $E_{\rm
cl}\approx 9$. The classical energy essentially increases
quadratically with $w_H$. On the other hand, the dependence on $w_P$
appears to be weak, even for sizable $a_P\ge0.5$. For moderate
$\xi_1\approx\pi/4$, the classical energy is again minimized by narrow
Higgs fields while the  gauge fields tend towards intermediate sizes
($w_G\approx3$). However, the  value for the (minimal) classical
energy at fixed $\xi_1\sim\pi/4$ is increased by about a factor 2 as compared to $\xi_1\sim0$. In the regime of its maximal 
value $\xi_1\sim \pi/2$, the (minimal) classical energy is even further
increased to about $30$. Again, in this case the minimal value occurs at 
non--zero extension of the Higgs field $w_H\approx3$. When we study the effect 
of the additional contribution from $f_P$, we also see that the
minimal classical  energy occurs at small $\xi_1$ and it does not vary
much in the range  $a_P\in[0.1,0.5]$. Beyond that regime the classical
energy starts rising  linearly with $a_P$.

We find it worth mentioning that the configuration we have investigated
in detail with the lowest classical energy does not have exactly
$\xi_1=0$, where the gauge field part in the background configuration
vanishes identically, but rather $\xi_1=0.05\pi$. As we have seen
before, even for this small value of $\xi_1$ the gauge field energy 
is of a noticeable size. We note, however, that $\xi_1=0$ is 
problematic for the numerical investigation as will be discussed 
further below, {\it cf.\@} footnote~9.

To some extent the situation is different for large fermion masses. 
Even though the classical energy does not explicitly depend on $m_f$,
this quantity enters via the VEV in $D=2+1$ dimensions,
eq.~(\ref{eq:vev3d}), and also affects the numerical result for
$E_{\rm cl}$ as it sets the overall energy scale\footnote{If one
re--expresses the classical energy by replacing couplings by
physical masses and the Fermi coupling and takes into account that
the energy is measured in units of $m_f$ it is obvious that the
Higgs field contribution to the energy is suppressed relative to
gauge field contribution when $m_f$ gets large.}. We have 
numerically studied the case with $m_f=1.5{\rm GeV}$. For small
$\xi_1$ the classical energy is also small, {\it i.e.\@} for
$\xi_1\le0.1$ we find $E_{\rm cl}\lesssim0.4$. 
The numerical values for the classical energy seem quite insensitive
to the shape of the Higgs field as the coupling via the VEV
is drastically decreased. Of course, for small $\xi_1$ the dependence
on $w_G$ is again mitigated. This changes for moderate values of
$\xi_1$. For $\xi_1\approx0.5$ we have varied $w_G\in[1.5,5.5]$.
In that range $E_{\rm cl}$ strongly varies between approximately
$2$ and $25$ with the smallest result obtained for the largest
width $w_G$. On the other hand, the dependences on the Higgs
field variational parameters is soft. We observe a similar
situation for $\xi_1$ in the vicinity of $\pi/2$, when the
relative contribution from the gauge fields is maximal.
However, the range along which $E_{\rm cl}$ changes is even
bigger. For small $w_G$ ($\sim1.5$), it may be as big
as $50$ while it reaches values close to one for very large
widths of the gauge fields $w_G\approx12$. We do not thoroughly study 
such configurations because in these cases the computation of the 
vacuum polarization energy is numerically very costly.
Minimization of the classical energy requires a
suppression of the gauge boson contribution (small $\xi_1$) and/or
a sizable extension of these fields. In the following sections we
will study the behavior of the fermion vacuum energy under these
conditions. In particular, we will study the question of whether
or not the energy gain due to the emergence of a fermion zero mode
at $\xi_1=\pi/2$ can compensate for the increase of the classical
energy.  The energy gain per occupied zero mode is unity and thus
we expect that we need a large number of fermions (twenty or so) to
break even, unless something unexpected happens to the vacuum
polarization energy.

As mentioned above, the classical energy does not vanish for the
limiting case of our variational parameters in which
$\xi_1$, $w_H$ and $a_P$ are all zero. The corresponding 
classical energy $E_{\rm cl}\sim6$ is the minimal value
that can be accommodated by our parameterization of the fields.
For this configuration, we expect the fermion spectrum to be that of
the trivial vacuum, hence there appears to be a minimal energy
threshold that must be overcome by finite size background fields.

\subsection{Bound States}

Central to our analysis are the bound states of the $W$--
and $Z$--strings.  For $\xi_1 = \frac{\pi}{2}$ and $\phi_+ = 0$, an
exact zero mode exists.   
Contrary to early expectations, it does not stabilize the $Z$--string
configuration against $\phi_+$ condensation once effects of the fermion
determinant are taken into account, as was shown in the perturbative
analysis of \cite{Naculich:1995cb}.
By taking the whole bound state spectrum into account we find
something unexpected. In the background of rather wide
strings (several Compton wave lengths of the fermion) we find a very
large number of bound states.  Because of this large number of
bound states (both spread over a lot of angular momentum channels and
in individual channels) the properties of individual states like the
zero mode are a lot less important than previously thought.  This
result provides an important reason for why the purely scalar
configurations are favored over the configurations containing gauge
fields. For the purely scalar configurations the lowest--lying 
states are less strongly bound, however, and contrary to expectations
from potential theory in quantum mechanics, many more bound 
states exist than for those configurations that generate strongly bound 
states. Thus the decrease of binding of the lowest--lying states
makes only a small destabilizing contribution. But the
price in classical energy to pay for gauge field configurations is
rather high, as we have already discussed. Hence it can be
advantageous to switch them off.

Before discussing the numerical results quantitatively we would
like to introduce criteria for their accuracy.  Numerically we find
the bound states via a shooting method, in which we determine the
number of bound states in each channel by Levinson's theorem. 
If our numerical precision is not high enough, the shooting method will
miss bound states that are very close to threshold\footnote{Another 
problem is that a shooting algorithm might miss states 
in pairs if they are degenerate in energy. Such a degeneracy actually
occurs for $\xi_1 = 0$ because the $CP$ symmetry is enhanced to an
individual $C$ and $P$ symmetry. We counter this problem by avoiding
the special case $\xi_1=0$.}.
Since each state contributes to the total energy only through its
binding energy, these states are relatively unimportant.  Therefore
it is more efficient to construct an accuracy criterion
from that observation and reject those computations that do not
satisfy that criterion.  Let $p_\ell$ be the number of missed bound
states and $\omega_\ell$ be the largest bound state energy that we
observed in the channel with angular momentum $\ell$.  From
eq.~(\ref{eq:Evac3}) an upper limit for the numerical error in the
vacuum energy is then
\be
\Delta E_{\rm vac}=\frac{1}{2}\sum_{\ell=-\infty}^\infty p_\ell
\left(m_f-\omega_\ell\right)\,,
\label{eq:numerr}
\ee
and we reject any computation for a given set of variational parameters
that exceeded $\Delta E_{\rm vac}$ by a prescribed value, which we
generally take as 5\% of the vacuum polarization energy, defined
in eq.~(\ref{eq:Evac3}) with $D_j=1$. 
Most of the rejected calculations are characterized by very small $\xi_1$. 
\begin{figure}[htbp]
\centerline{~~\hskip1cm
\includegraphics[height=9.6cm,width=6.5cm,angle=270]{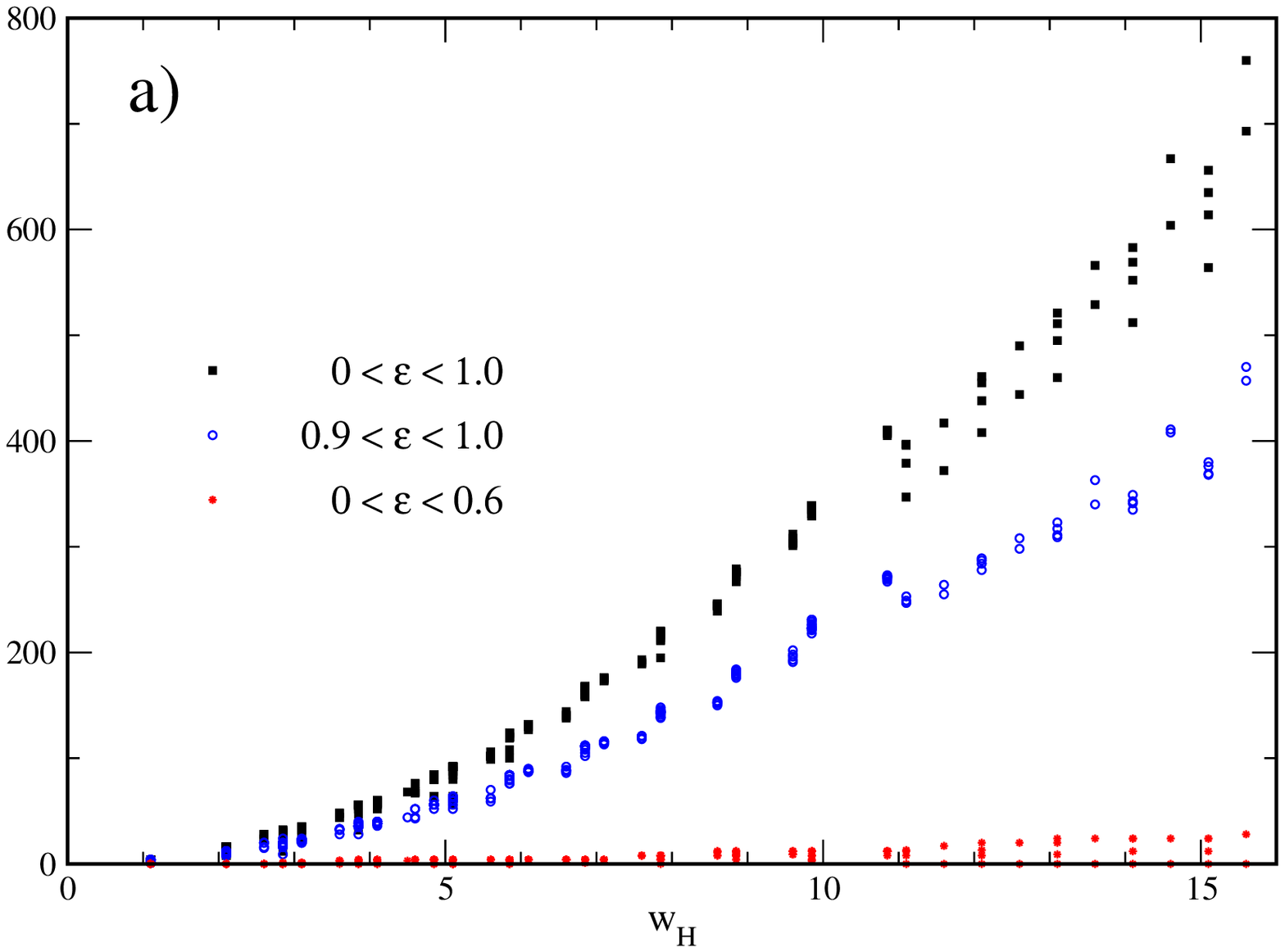}~~
\includegraphics[height=9.6cm,width=6.5cm,angle=270]{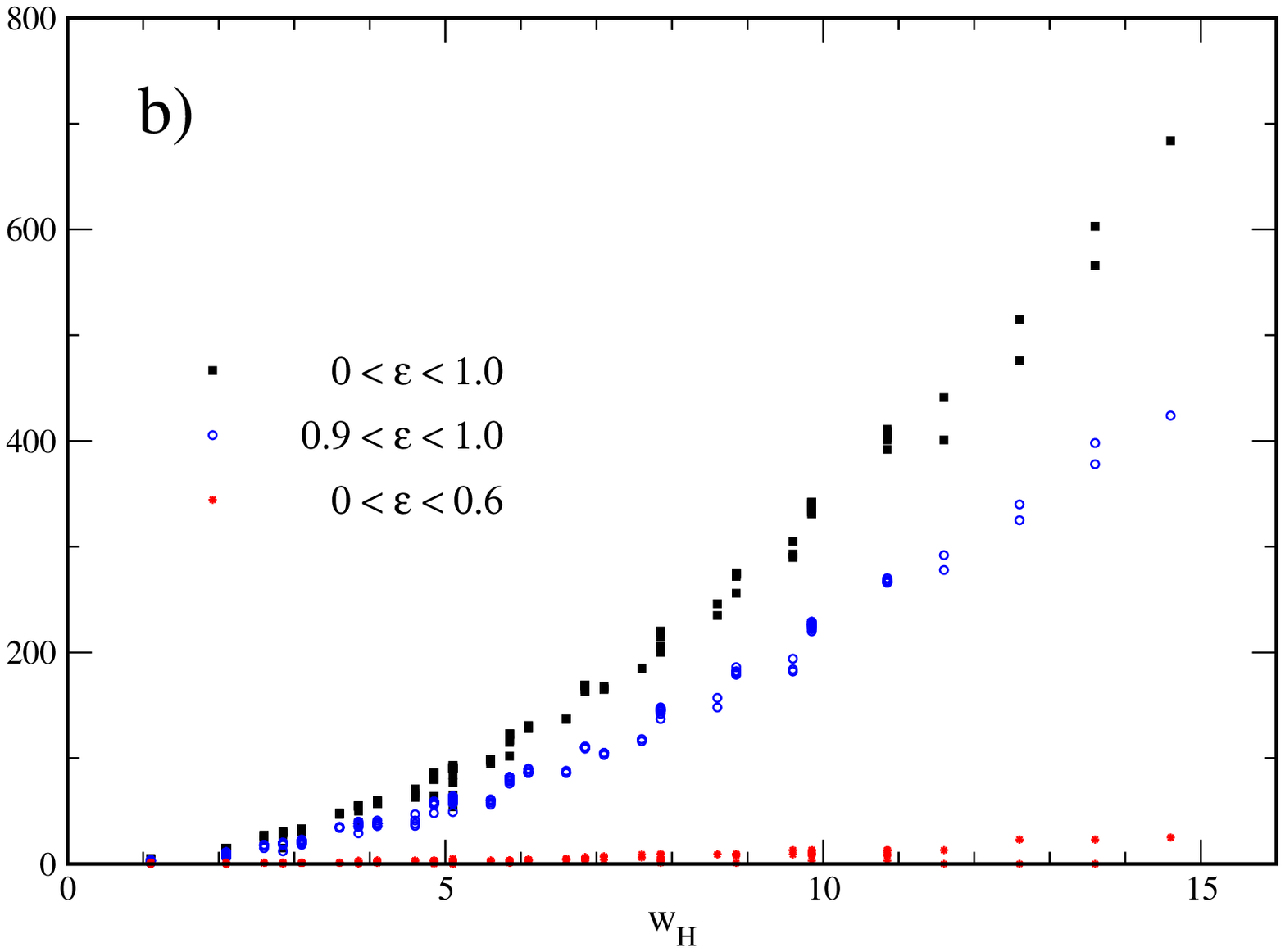}}
\caption{\label{fig:bsnum}The number of bound states as function of
the width $w_H$ of the Higgs field for two different values of the
mixing angle: $\xi_1=0.1 \times \pi/2$ (a) and $\xi_1=\pi/2$ (b).
We distinguish between strongly ($0<\epsilon<0.6$) and loosely
($0.9<\epsilon<1.0$) bound states. The entry $0<\epsilon<1.0$
obviously gives the total number of bound states. The towers in each
entry show results that originate from various values of those variational
parameters that are neither held fixed nor are shown along the abscissa.
}
\end{figure}

In  figure~\ref{fig:bsnum} we display the number of bound states as a
function of the width parameter $w_H$. There are $n_\ell$ bound states in 
each orbital momentum channel and we display $\sum_\ell n_\ell$. First of
all we see that the number of bound states strongly increases roughly
quadratically with the width of the Higgs field. 
The number of bound states increases only gradually with 
increased coupling to the gauge fields as measured by $\xi_1$, because the 
main reason for the appearance of these bound states is that a 
Higgs field with $|\Phi|<|\langle \Phi \rangle|$ produces an attractive 
potential for the fermions, since $f|\Phi|$ is the effective fermion mass.
The number of strongly bound states also increases only gradually with 
$\xi_1$, which is somewhat surprising since we have a zero mode
for $\xi_1=\pi/2$ and $a_P=0$.\footnote{We did not consider a configuration
with $\xi_1=\pi/2$, $w_H=14.8$ and small $a_P$. Thus the $w_H=14.8$
entry in figure~\ref{fig:bsnum}b) misses a strongly bound
state.}  However, it is not generally accompanied by additional strongly bound
states.

\begin{figure}[htbp]
\centerline{~~\hskip1cm
\includegraphics[height=9.6cm,width=6.5cm,angle=270]{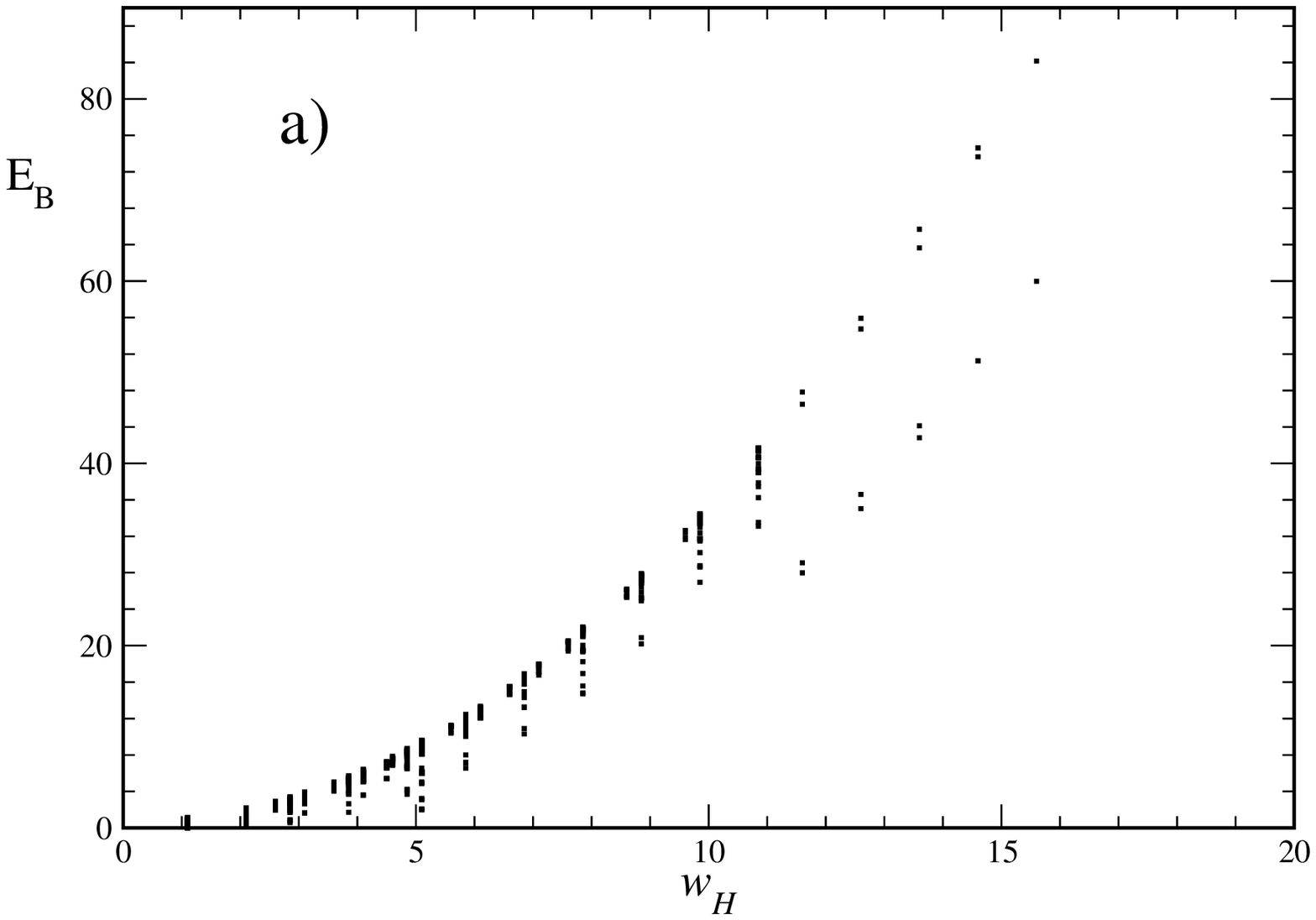}~
\includegraphics[height=9.6cm,width=6.5cm,angle=270]{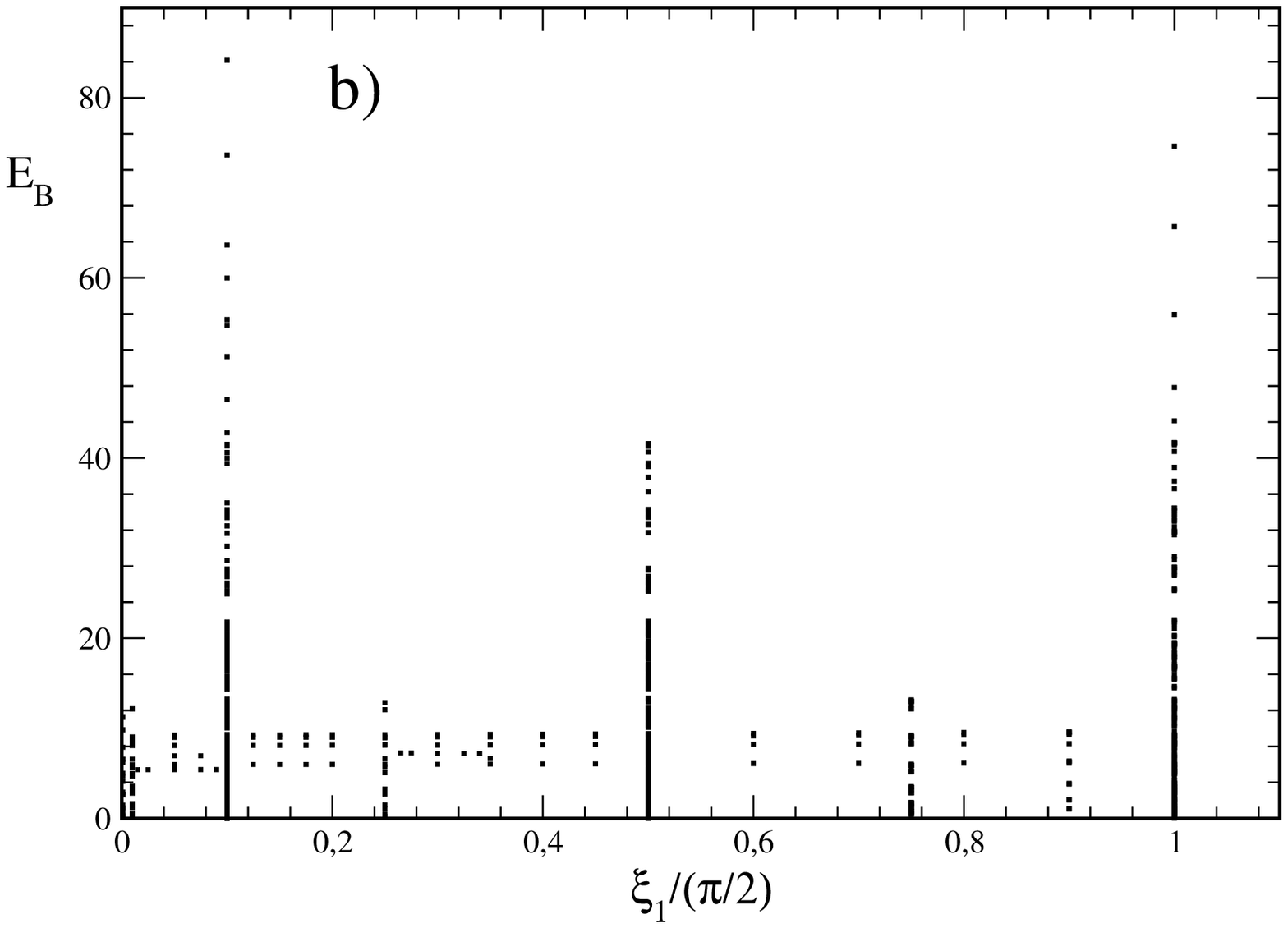}}
\caption{\label{fig:eball}The bound state contribution, $E_B$ to
the vacuum polarization energy as a function of the variational
parameters a) $w_H$ and b) $\xi_1$ that are
defined in eqs.~(\ref{eq:SphaleronSquare}) and~(\ref{eq:profiles}).
The towers in each entry show results that originate from various
values of those variational parameters that are not shown along
the abscissa. It should be emphasized here - especially with regards 
to panel b) - that our coverage of parameter space has not been uniform. 
For a large range of widths, e.g., computations have only been performed 
for $\xi_1 = 0.1, 0.5, 0.9$. Hence one may deduce from panel b)
that the range of energies is very similar for $\xi_1 = 0.1$ and
$\xi_1=0.9$, but not that the range of energy values at $\xi_1 = 0.2$ is
much smaller than for $\xi_1 = 0.1$ (in units of $\pi/2$).}
\end{figure}

This bound state structure suggests that the regime with small
coupling to the gauge bosons ($\xi_1\sim0$) might be of particular
interest in our analysis.  We consider the bound state contribution to
the vacuum energy
\be
E_{\rm B}=\frac{1}{2}\sum_{\rm b.s.}\left(m_f-|\epsilon_i|\right)
\label{eq:defEB}
\ee
and in figure~\ref{fig:eball} we display $E_{\rm B}$ as function of
$w_H$ and $\xi_1$.  We see that $E_B$ increases with $w_H$
approximately quadratically, while the dependence of $E_B$ on the
other variational parameters is weaker.  As an example we also show
$E_B$ as a function of $\xi_1$ in figure~\ref{fig:eball}b).
Even though $E_B$ is large for some of the configurations in the
vicinity of $\xi=\frac{\pi}{2}$ the maximal values are found for
small $\xi_1$. Since the classical energy grows with $\xi_1$,
it is unlikely to find stable objects in the region of large 
$\xi_1$, {\it i.e.\@} for $W$ dominated string
configurations; rather pure Higgs configurations seem to be favored.

\subsection{Vacuum Polarization Energy}

\begin{figure}[htbp]
\centerline{
\includegraphics[height=9.6cm,width=6.5cm,angle=270]{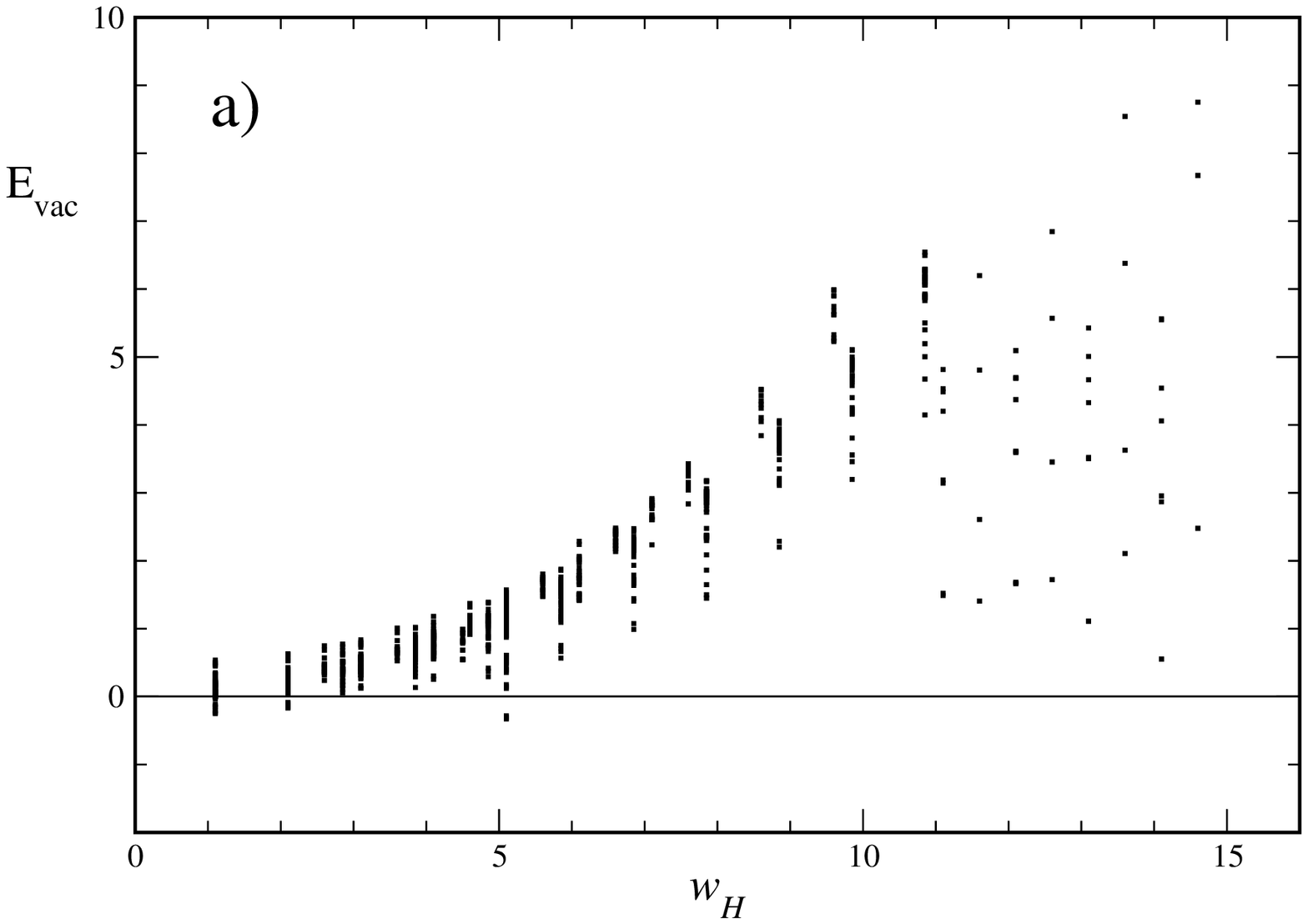}~~
\includegraphics[height=9.6cm,width=6.5cm,angle=270]{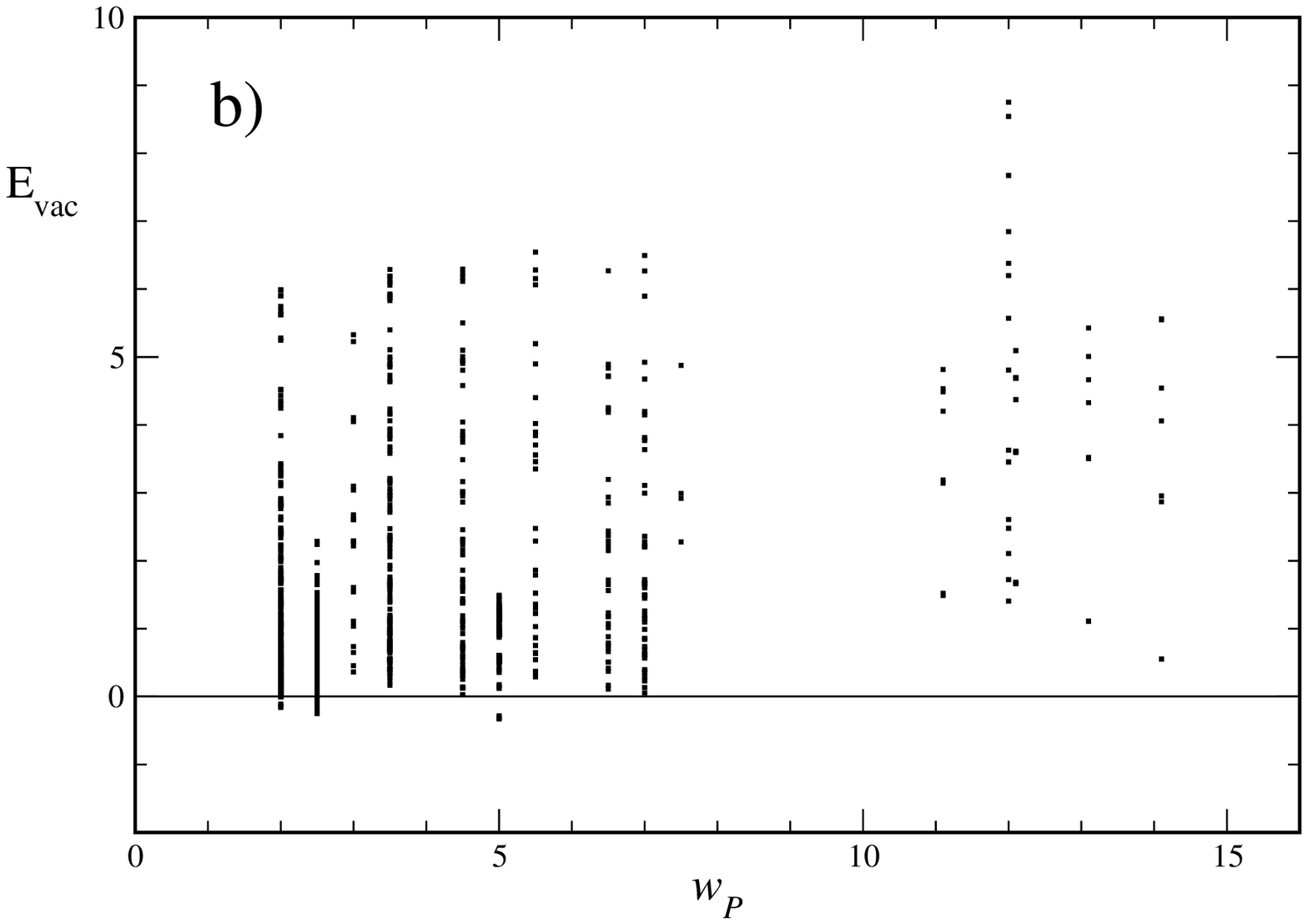}}
\centerline{
\includegraphics[height=9.6cm,width=6.5cm,angle=270]{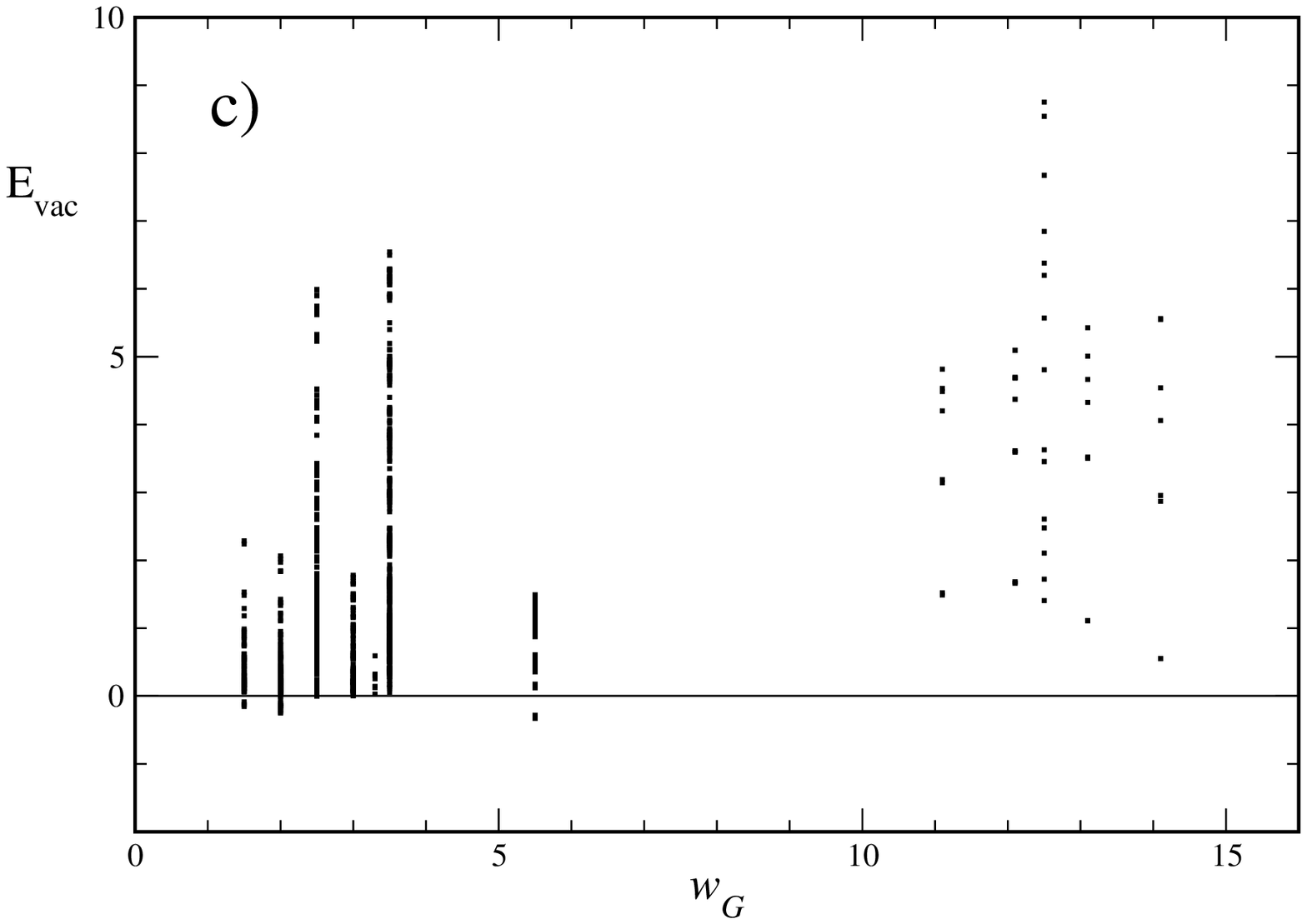}~~
\includegraphics[height=9.6cm,width=6.5cm,angle=270]{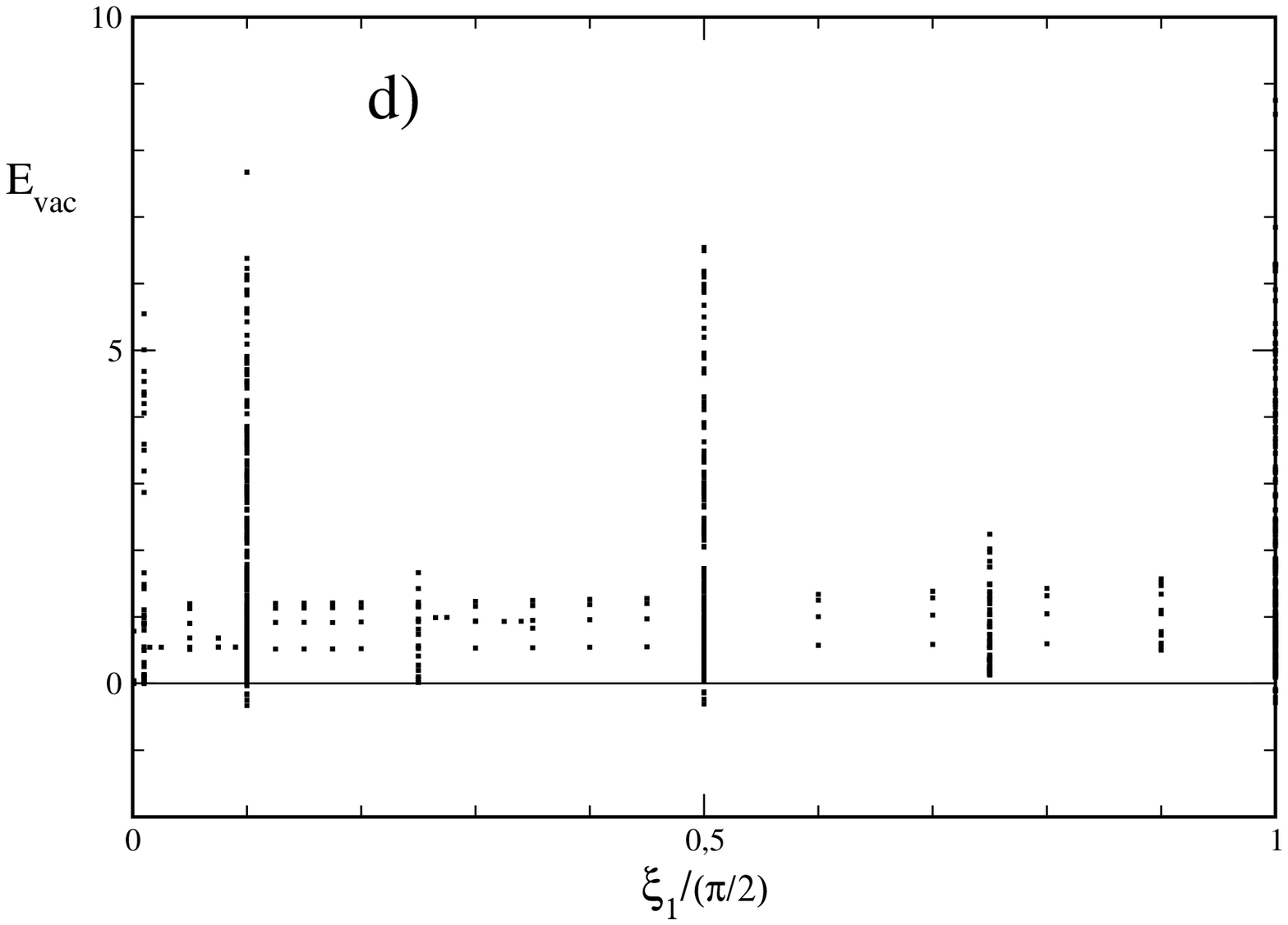}}
\centerline{
\includegraphics[height=9.6cm,width=6.5cm,angle=270]{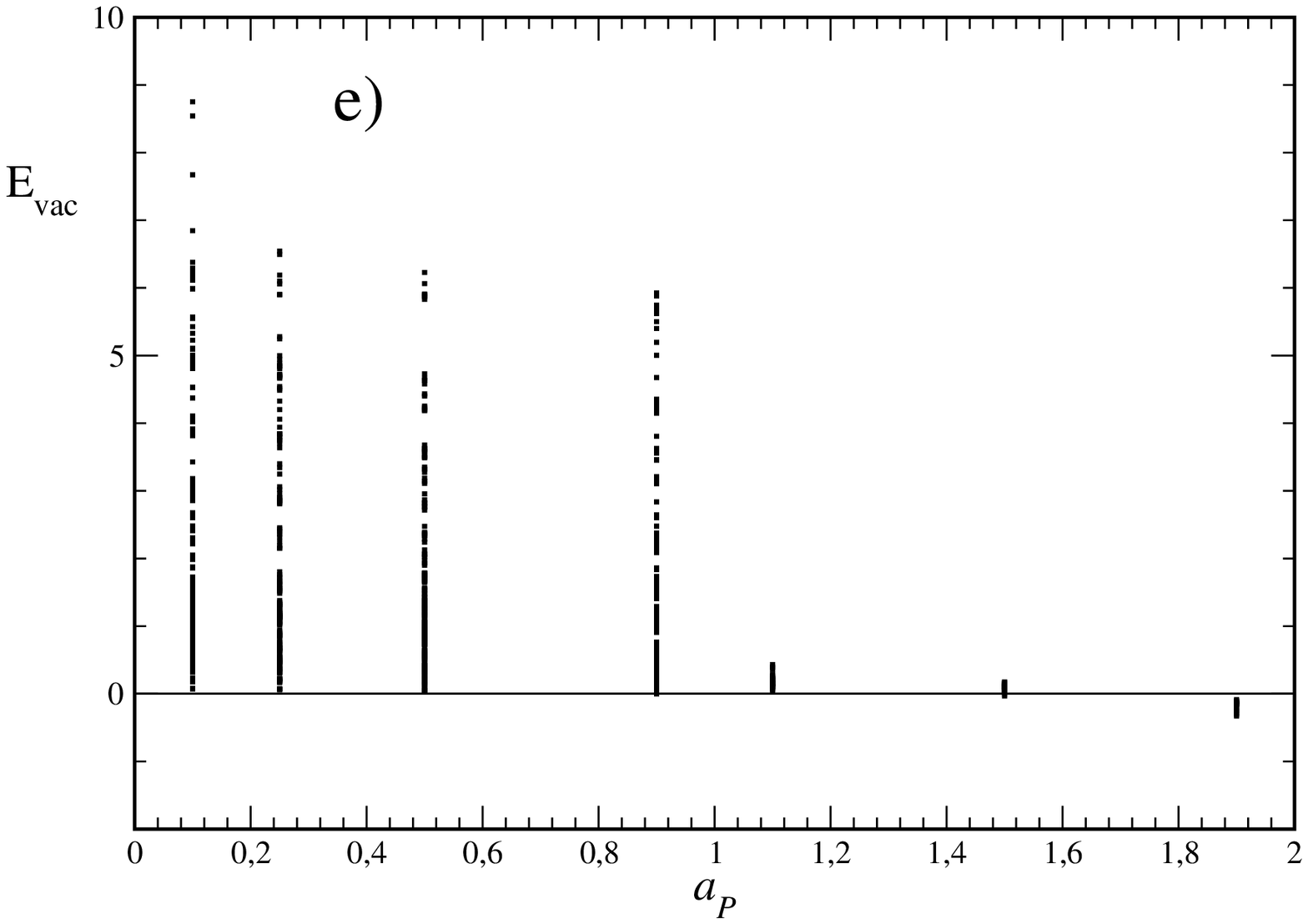}~~
\parbox[htbp]{7.4cm}{\caption{\label{fig:evac}The vacuum
polarization energy as a function of the variational parameters
a) $w_H$, b) $w_P$, c) $w_G$, d) $\xi_1$ and e) $a_P$ that are
defined in eqs.~(\ref{eq:SphaleronSquare}) and~(\ref{eq:profiles}).
The towers in each entry show results that originate from various
values of those variational parameters that are not shown along
the abscissa. With regard to panel a) it should be noted that the 
large spread of values for larger values of $w_H$ is due to finite 
accuracy: the vacuum polarization energy is computed by adding up several
components, displayed,e.g., in fig. \ref{fig:eder},  that each have a
finite accuracy. For large values of $w_H$ the numerical uncertainty
is of the order of the vacuum polarization energy; this of course
contributes to the spread making it larger.
For further comments, {\it cf.\@} fig~\ref{fig:eball}.}}
\parbox[htbp]{1.8cm}{~}}
\end{figure}

We now turn to the discussion of numerical results for the vacuum
polarization energy, $E_{\rm vac}$, which we compute according to
eq.~(\ref{eq:Evac3}). We have accumulated data for about 1000 sets
of variational parameters.  Our results are displayed in
figure~\ref{fig:evac} for the physical value of the top quark
mass $m_f=170{\rm GeV}$.

We note for some configurations the renormalized vacuum energy is
negative and can provide some energy gain to the system. The maximal gain 
is achieved for moderate extensions 
of the Higgs field $w_H,w_P\sim 3\ldots5$ while $E_{\rm vac}$
quickly increases when these extensions grow even further.  As
figure~\ref{fig:evac}c) indicates, the most favored value of the gauge
boson field extension is about $w_G\sim3$.  Both of these regions
could also lead to a small classical energy. On the other hand we do
not observe a strong dependence of $E_{\rm vac}$ on the angle $\xi_1$ {\it cf.\@} figure~\ref{fig:evac}d). That is, the vacuum polarization
part of the energy is \emph{not} capable of overcoming the strong
drift towards the Higgs dominated configuration ($\xi_1=0$)
caused by $E_{\rm cl}$.  In addition, we observe that minimization of
the vacuum energy induces a charged Higgs field, {\it i.e.\@}
$E_{\rm vac}$ is minimized by $a_P\ne$ as shown in
figure~\ref{fig:evac}e).

In all we find that the energy gain from $E_{\rm vac}$ is
only of order $-m_f$. The lowest value that we observed is
$E_{\rm vac}=-0.33m_f$ for $w_H=5.1$, $w_P=5.0$, $w_G=5.5$,
$a_P=1.9$ and $\xi_1=0.1\times(\pi/2)$, which is essentially a
Higgs background configuration.  So this energy gain alone is not sufficient
to stabilize the string, but we can still populate fermion levels
along the string to construct a stable object.

\subsection{Gradient Expansion}

\begin{figure}[htbp]
\centerline{~~\hskip1cm
\includegraphics[height=9.6cm,width=6.5cm,angle=270]{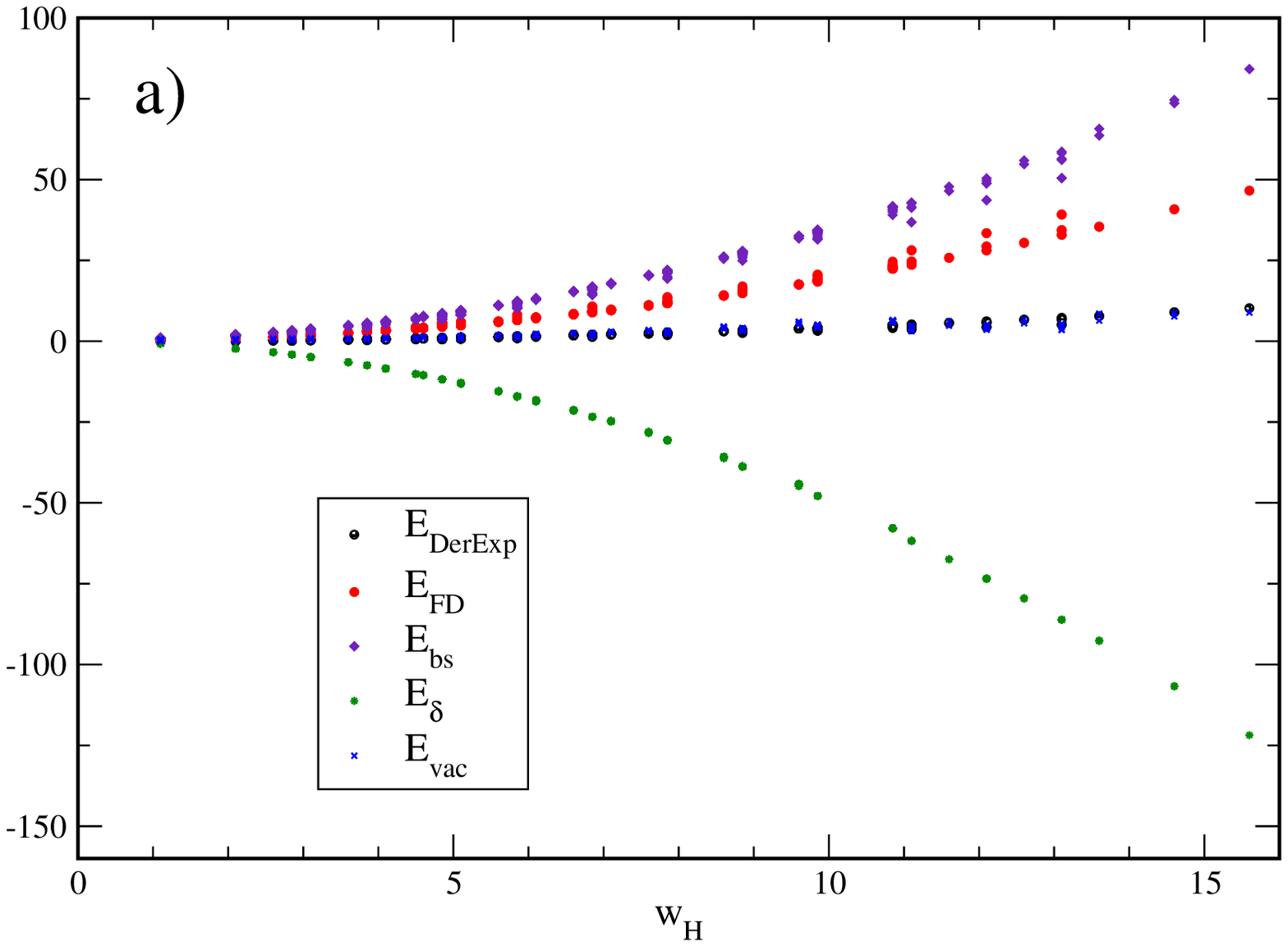}~~
\includegraphics[height=9.6cm,width=6.5cm,angle=270]{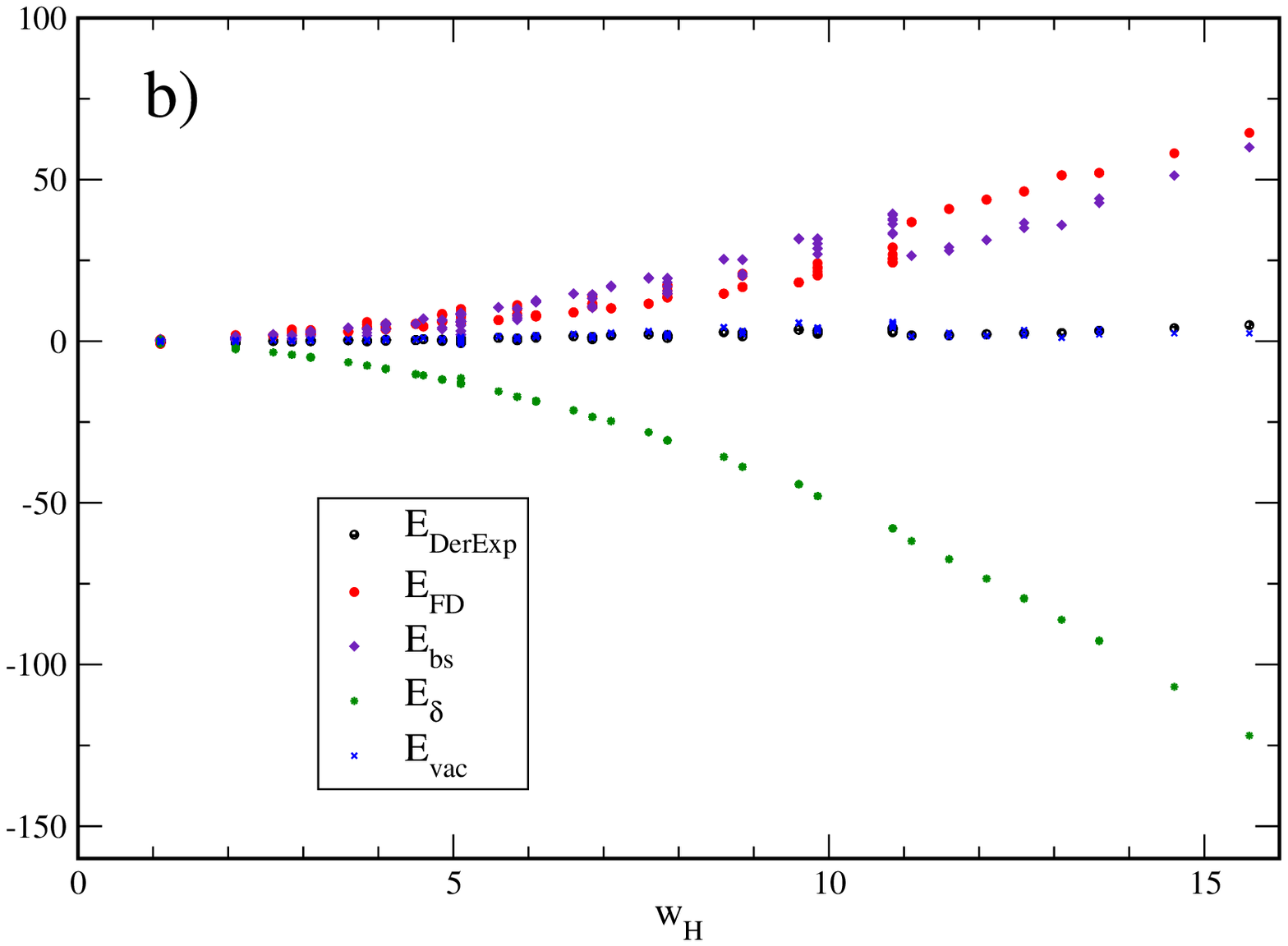}}
\caption{\label{fig:eder}Comparison of the gradient expansion
results ($E_{\rm DerExp}$, eq.~(\ref{eq:Evacapprox})) 
with the full calculation ($E_{\rm vac}$, eq.~(\ref{eq:Evacgeneral})).
We also disentangle the various contributions to $E_{\rm vac}$.
The results are shown as functions of $w_H$, and we furthermore
split the remaining parameters into sets with $a_P<0.8$ (a) and
$a_P\ge0.8$ (b).}
\end{figure}

We have already introduced the gradient (or derivative)
expansion approximation as a method of checking our numerical results.
In figure~\ref{fig:eder} we display the comparison of the
full result, $E_{\rm vac}$, which we compute according to
eq.~(\ref{eq:Evacgeneral}), and the gradient expansion
approximation, as discussed in eq.~(\ref{eq:Evacapprox}).
We display these data at the same scale
as the various pieces that contribute to $E_{\rm vac}$
to show that the sizable cancellations that occur among
these piece are necessary to accomplish the agreement with
the gradient expansion approximation.
In particular the case $a_P\ge0.8$ shows that changes in
the bound state contributions $E_{\rm bs}$ are compensated
by a corresponding change in the (renormalized) Feynman diagram
contribution, $E_{\rm FD}$. Na{\"\i}vely we expect compensations
between the bound  state and phase shift contributions as a reflection
of Levinson's theorem. However, the Born subtractions relocate
substantial pieces of the  phase shift contribution into the Feynman
diagrams. We only show the results as functions of the width of the
Higgs field, $w_H$, because the dependence of the 
subtracted phase shift contribution $E_\delta$ on variational parameters
other than $w_H$ is negligible. Some small discrepancy between
the $E_{\rm vac}$ and $E_{\rm DerExp}$ occurs at small $\xi_1$
because of the numerical problems we encounter in finding very weakly
bound states in that case.  However, this is not too  worrisome
because we can easily extrapolate from the region $\xi_1\ge0.01$.

\subsection{Stabilization with Populating Levels}

We have seen that many fermion bound states emerge 
for the background string configuration. We may populate
these levels to describe an object with fermion number
$N_f N_C$ and effective energy 
\be
E_{\rm eff}^{(N_f)}=N_C\sum_{i=1}^{N_F}\epsilon_i+ N_C E_{\rm vac}
+E_{\rm cl}\,,
\label{eq:effnf}
\ee
where the sum over the bound states is such the $N_f$ lowest
energy modes are included. The corresponding binding energy is
\be
B^{(N_f)}=E_{\rm eff}^{(N_f)}-N_C N_f m_f
=N_C\sum_{i=1}^{N_f}\left(\epsilon_i-m_f\right)+ N_C E_{\rm vac}
+E_{\rm cl}\,.
\label{eq:Bf}
\ee
Obviously $B^{(N_f)}<0$ indicates an energetically stable object with
fermion number $N_f N_C$. If there are fewer than $N_f$ bound
levels for the configuration under consideration, we have to ``occupy''
scattering states at threshold, which do not contribute to the
binding energy.
\begin{figure}[htbp]
\centerline{~~\hskip1cm
\includegraphics[height=12.0cm,width=6.0cm,angle=270]{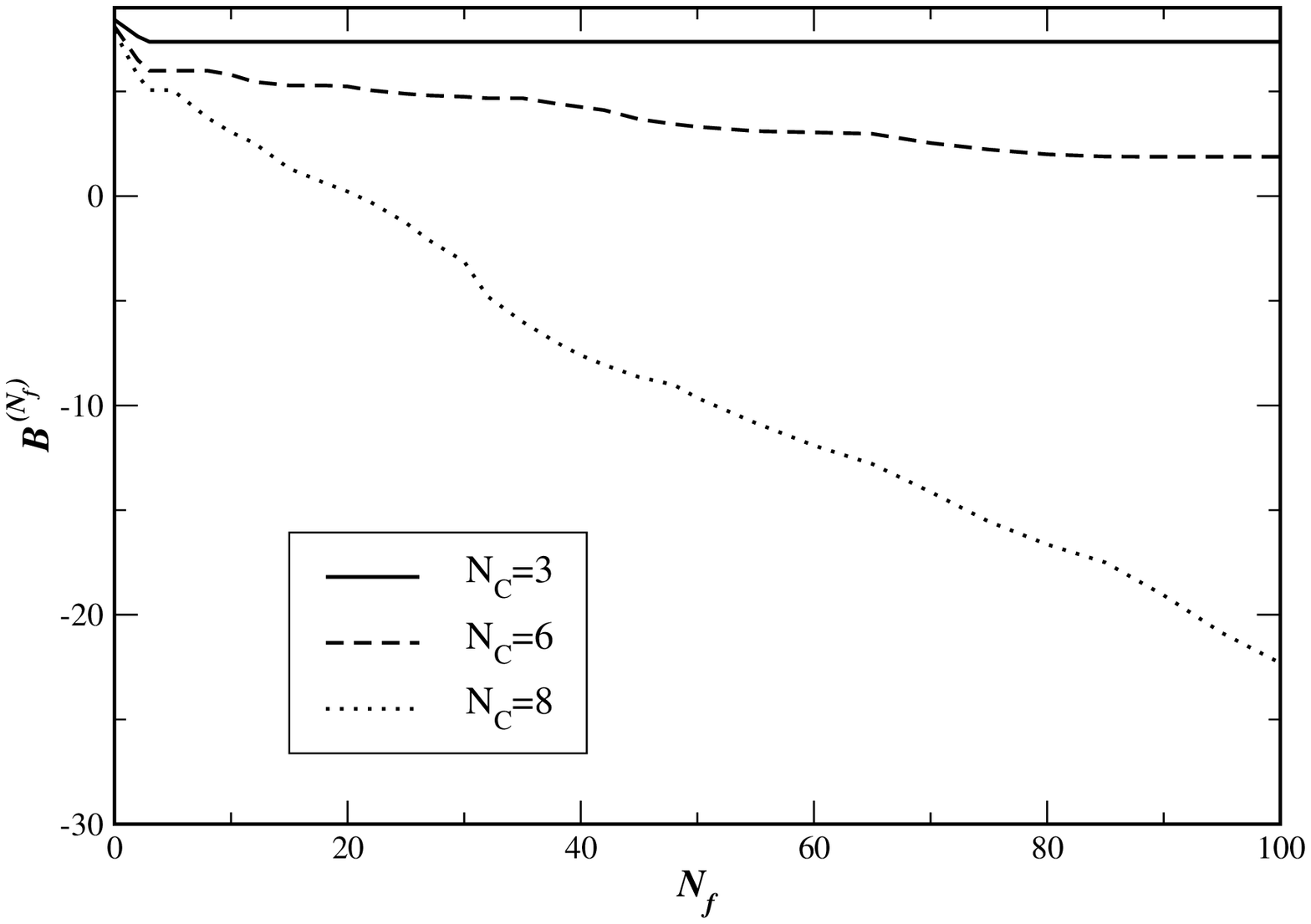}}
\caption{\label{fig:effnf}The extremal (negative) binding energy,
$B^{(N_f)}$ as a function of $N_f$, the number of populated bound
state levels.}
\end{figure}
In figure~\ref{fig:effnf} we display the extremal value of the
binding energy for a given number ($N_f$) of populated levels.
That is, for a prescribed value of $N_f$ we extract the lowest
$B^{(N_f)}$ in the space of variational parameters.  Bound 
objects exist for sufficiently large $N_C$.  In this case $N_C\ge8$ 
yields a bound object with $N_f>20$.

For small $N_C$, the effective energy is dominated by $E_{\rm cl}$,
which in turn is minimized by small $w_H$. In that case there
are only few bound states to be populated and $B^{(N_f)}$ is
saturated once all of them are occupied.  For larger $N_C$, 
the fermion contribution to the effective energy becomes more
important and the minimal value is obtained at larger widths,
where there are more bound states and the saturation of $B^{(N_f)}$
sets in at larger values of $N_f$.  This behavior is also reflected in
figure~\ref{fig:effnf}.

\begin{figure}[htbp]
\centerline{~~\hskip1cm
\includegraphics[height=9.6cm,width=6.0cm,angle=270]{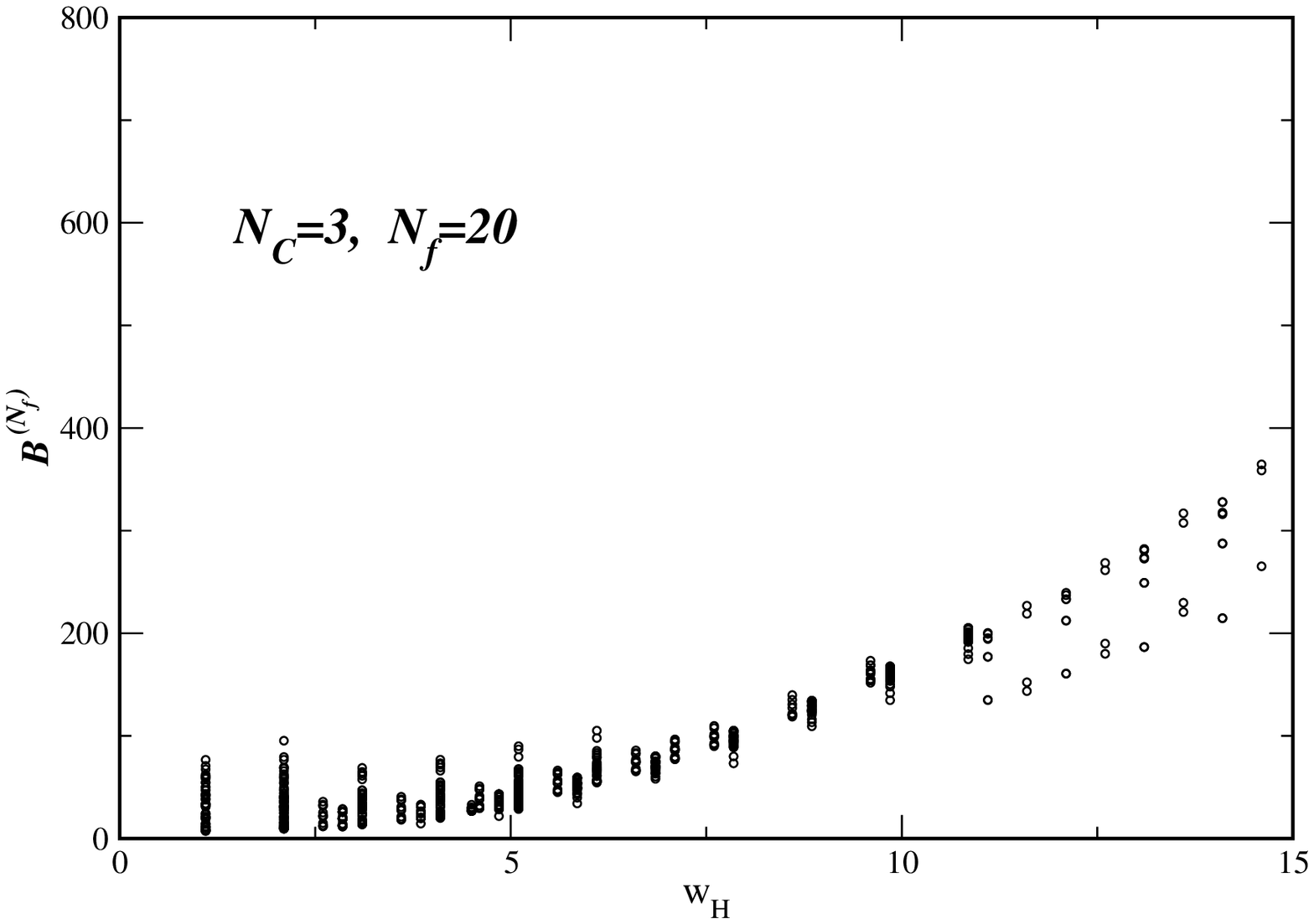}~~
\includegraphics[height=9.6cm,width=6.0cm,angle=270]{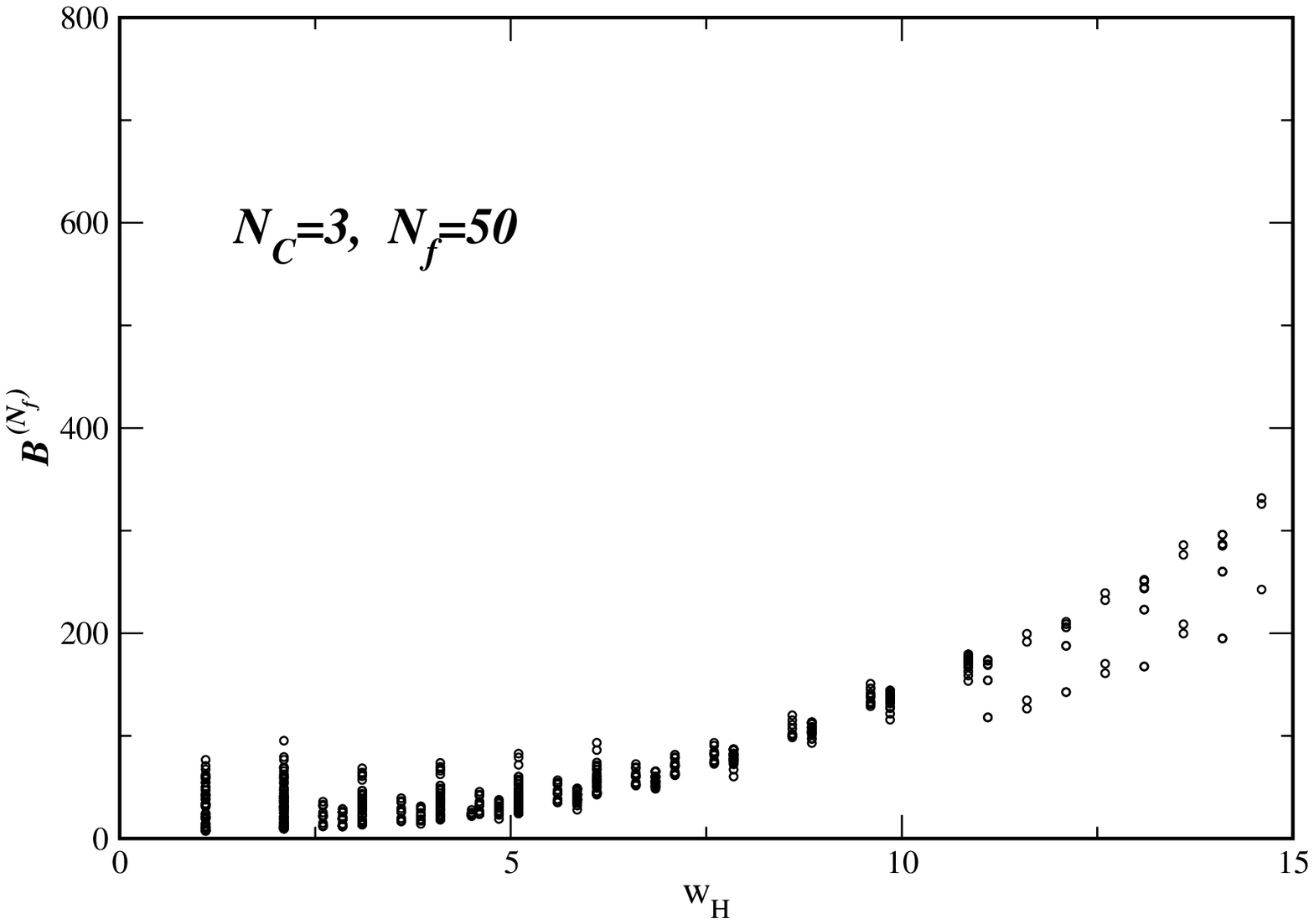}}
\centerline{~~\hskip1cm
\includegraphics[height=9.6cm,width=6.0cm,angle=270]{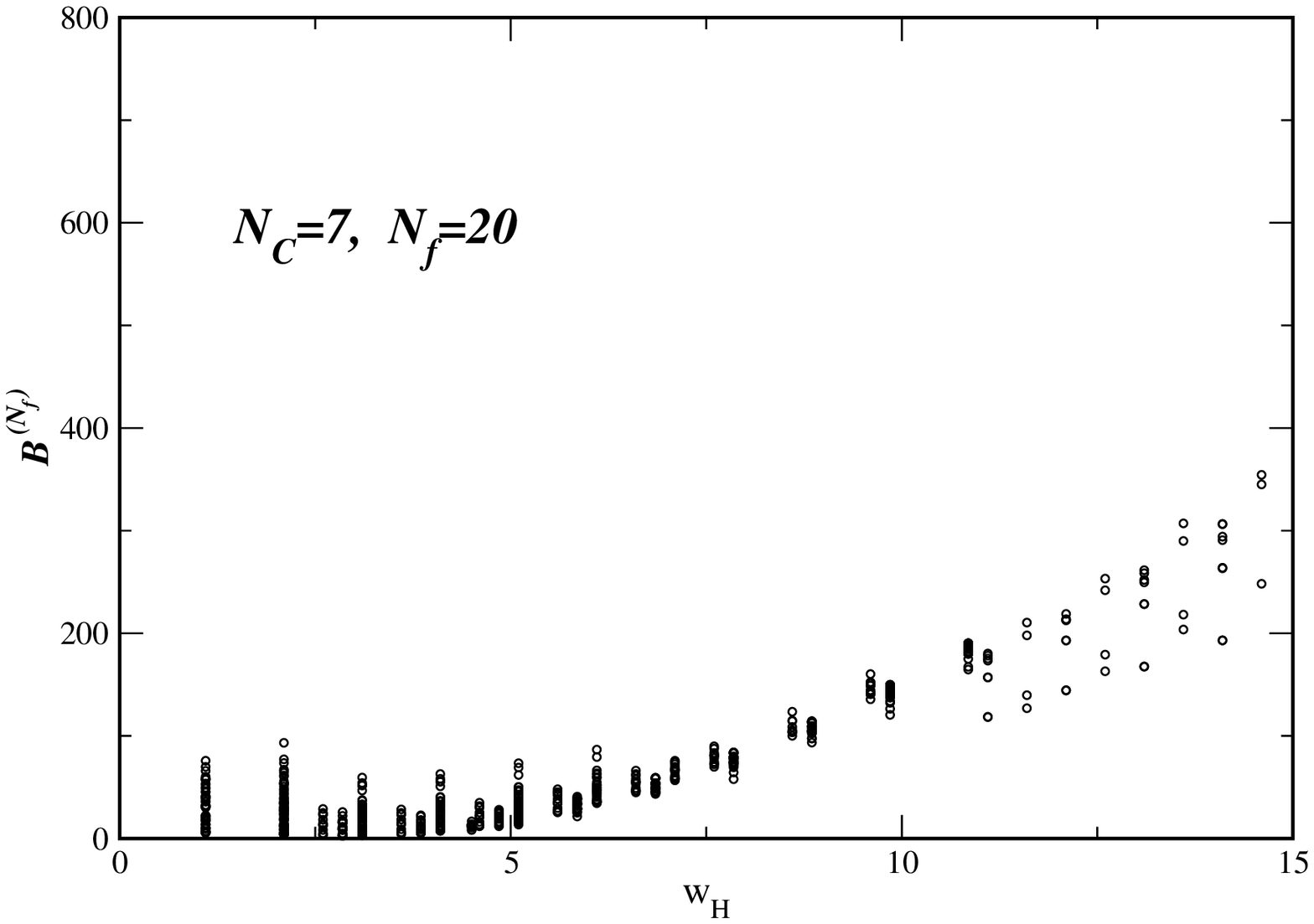}~~
\includegraphics[height=9.6cm,width=6.0cm,angle=270]{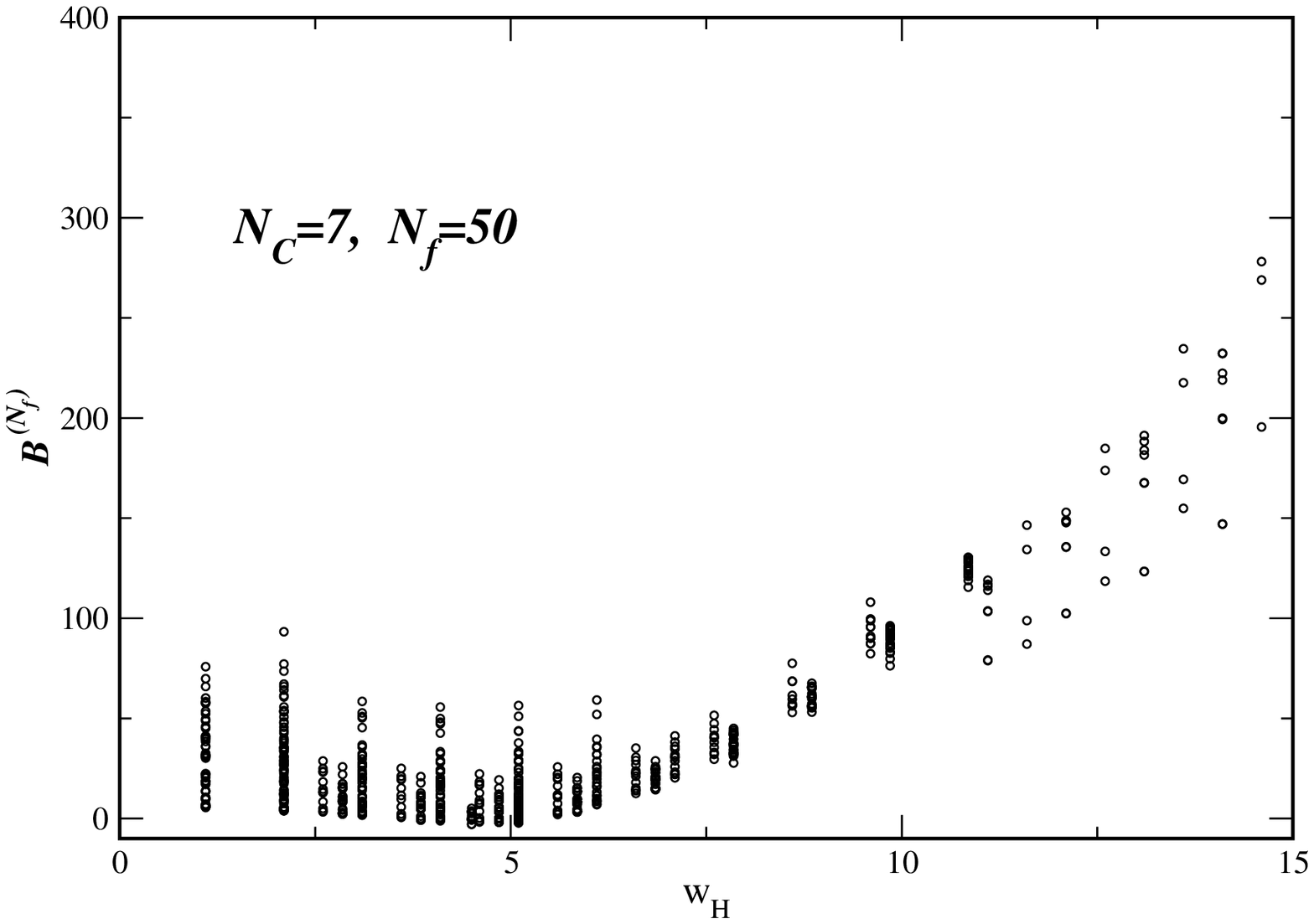}}
\caption{\label{fig:ebncnf}The binding energy, eq.~(\ref{eq:Bf}) as
function of the width parameter $w_H$ for various values of
$N_C$ and $N_f$.}
\end{figure}

In figure~\ref{fig:ebncnf} we display the binding energy
eq.~(\ref{eq:Bf}), for prescribed values of $N_C$ and $N_f$, as a
function of $w_H$ to demonstrate that we indeed obtain
a local minimum in the energy when the total fermion number is constrained.
Again, we see that a bound object requires a large value for the
number of colors.

\begin{figure}[htbp]
\centerline{~~\hskip1cm
\includegraphics[height=9.6cm,width=6.0cm,angle=270]{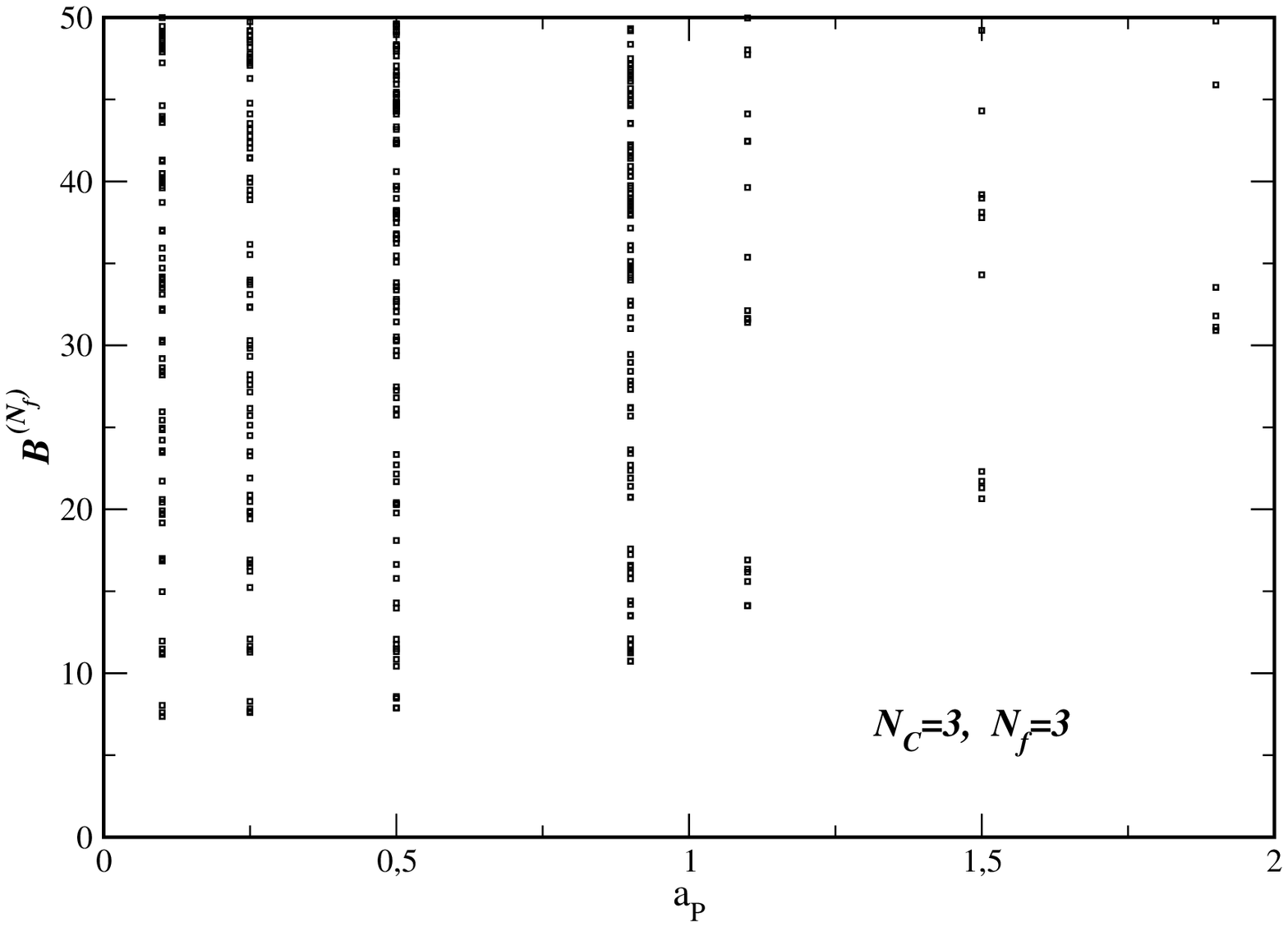}~~
\includegraphics[height=9.6cm,width=6.0cm,angle=270]{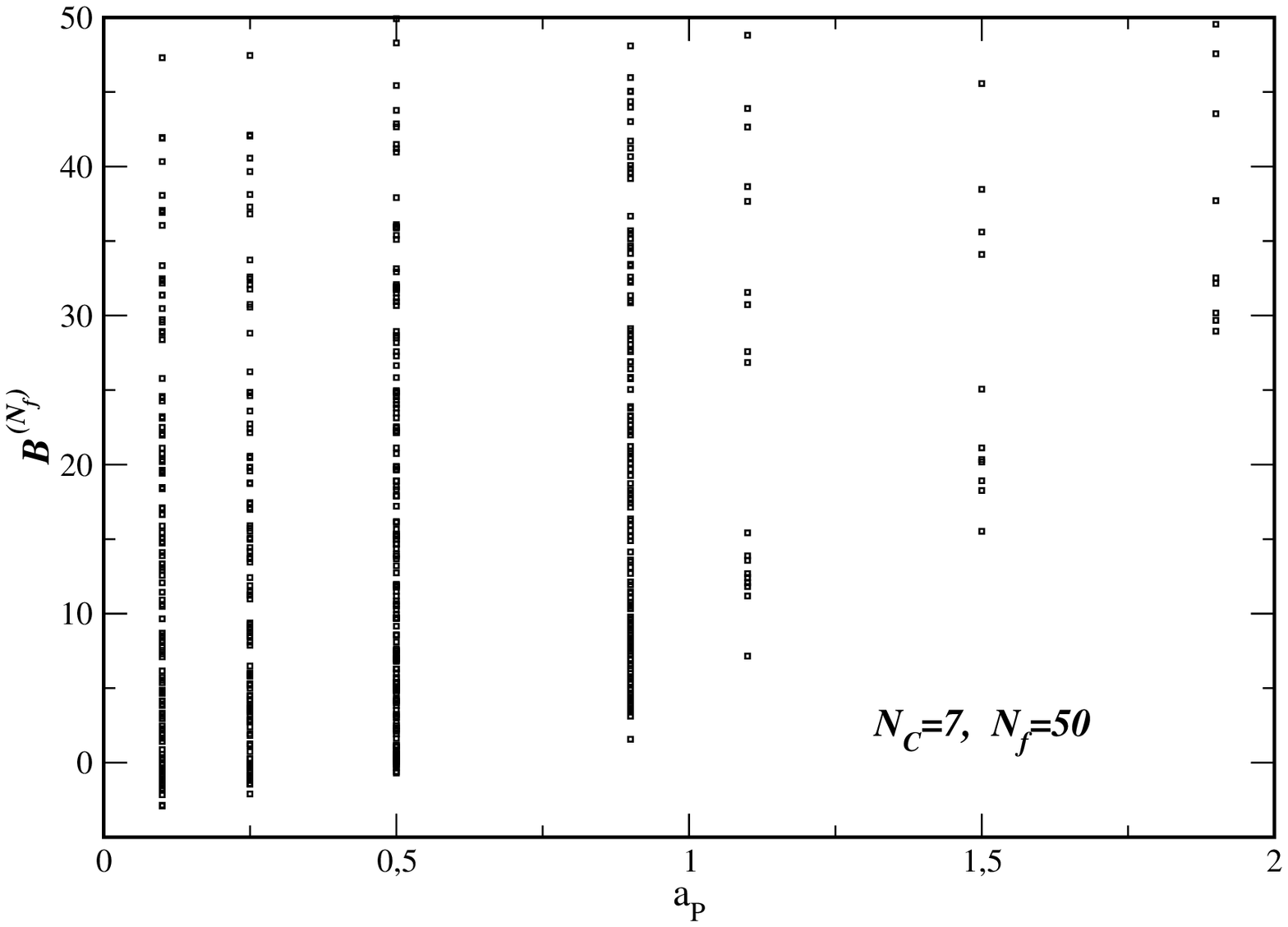}}
\caption{\label{fig:eap}The binding energy as function of the
amplitude of the charged scalar field $a_P$ for two values
of the number of populated levels.}
\end{figure}

The additional variational term $a_P$ in eq.~(\ref{eq:SphaleronSquare})
is novel in  the context of strings in the standard model and it
is worthwhile to briefly reflect on its impact on the fermion vacuum
energy.   The fermion zero mode exists for $\xi_1=\frac{\pi}{2}$ only
for $a_P=0$.  In addition any non--zero value of $a_P$ increases the
effective fermion mass.  Thus we expect largest binding in the small $a_P$
regime.  Indeed we find that $B^{(N_f)}$ is maximal for $a_P\to0$. It is
thus a bit surprising that $a_P$ actually decreases the vacuum energy,
{\it cf.} figure~\ref{fig:evac}b); $E_{\rm vac}$ may even become
negative only for sufficiently large $a_P$. As a consequence, at least
for objects with moderate fermion numbers, the total energy does not
increase with $a_P$ despite that the energy eigenvalues of the individual
fermion modes get larger. This effect is shown in figure~\ref{fig:eap}.

However, for these moderate values of $N_f$ overall binding is not yet
observed.  When we turn to the cases that exhibit
binding, the system prefers $a_P\to0$ because the bound states
become more important as we turn up $N_f$. This effect is shown in
figure~\ref{fig:eap}.

\subsection{Large Fermion Mass}

We have seen that stable objects can be constructed by choosing
unconventional parameters. In particular, increasing the number
of fermion degrees of freedom helps to bind objects with large fermion
numbers, since it suppresses the classical energy in comparison to the
fermion determinant and bound state contributions.  The obvious
question is whether there are other regions in the space of model
parameters that allow for stable objects.  Such a search, of course,
cannot be exhaustive and we will therefore just report on a single
calculation in that direction. The most obvious change in parameters
that enhances the role of the fermion fluctuations is an increase in
the Yukawa coupling, $f$.  This change increases the fermion mass and
simultaneous decreases the VEV of the Higgs field in $D=2+1$ due to
our definition eq.~(\ref{eq:vev3d}).  Since $v=v_{D=2+1}$ sets the
scale of the classical energy, this change represents an additional
enhancement of the fermion vacuum contribution.

\begin{figure}[htbp]
\centerline{~~\hskip1cm
\includegraphics[height=9.6cm,width=6.0cm,angle=270]{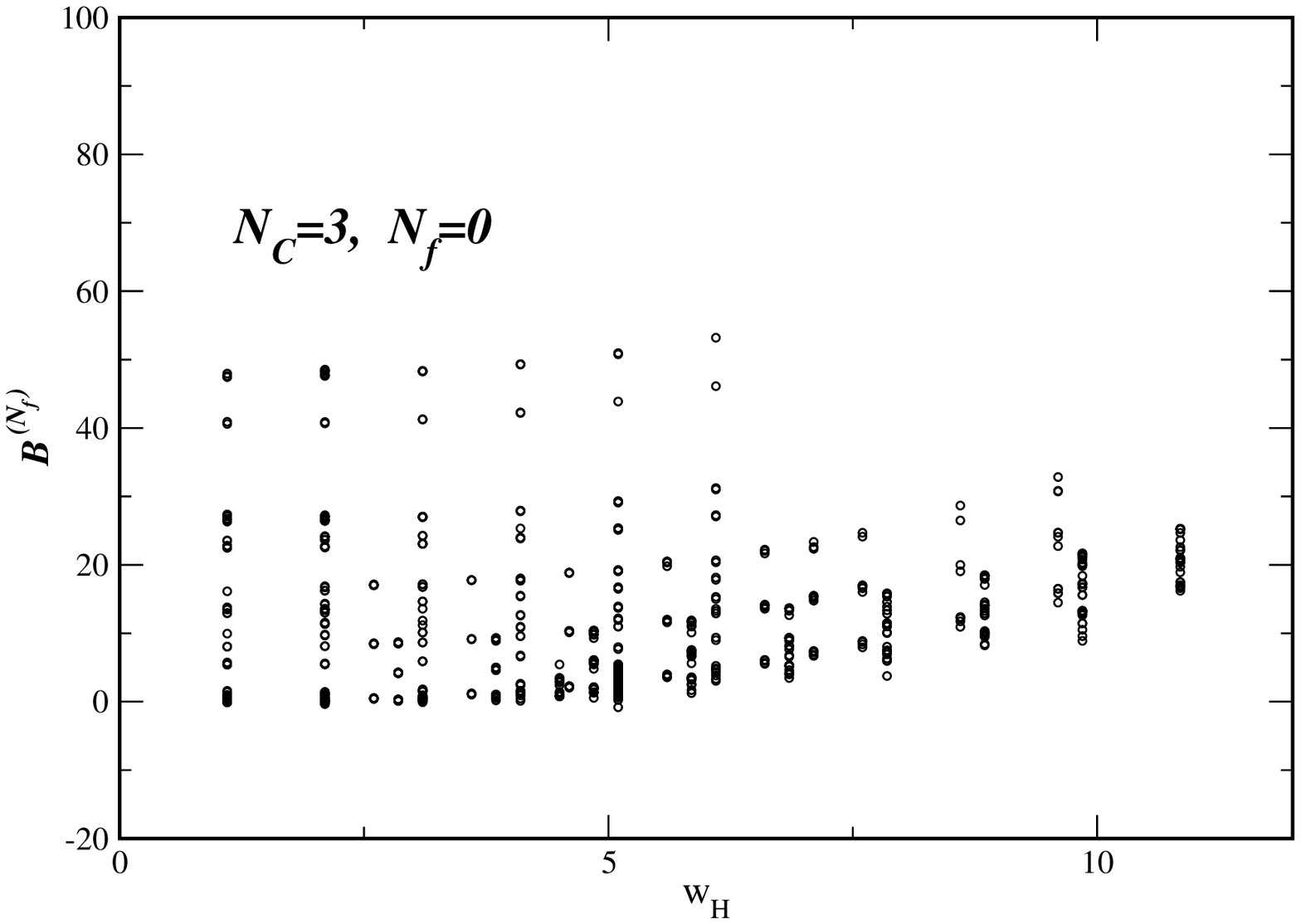}~~
\includegraphics[height=9.6cm,width=6.0cm,angle=270]{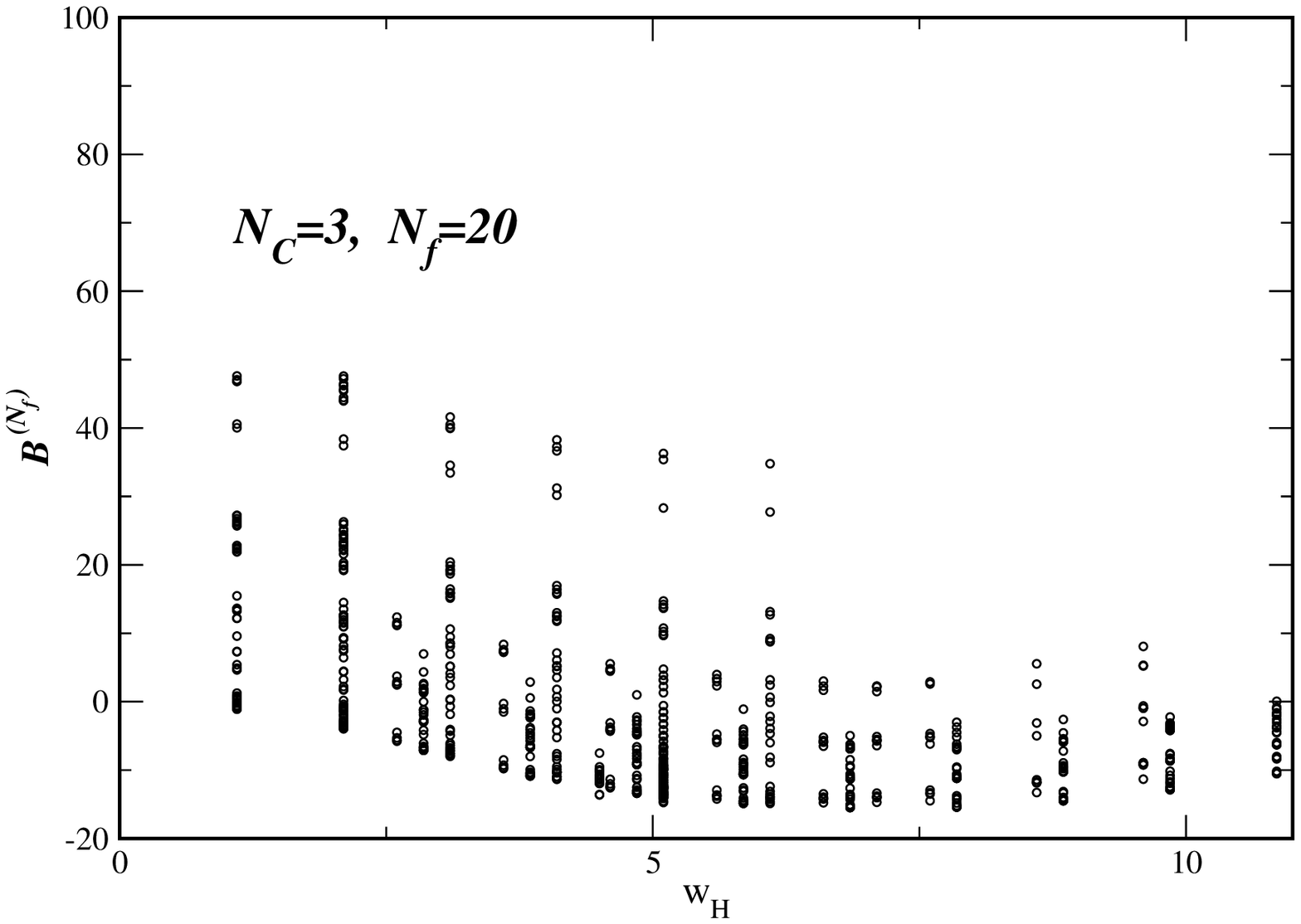}}
\caption{\label{fig:emf}The binding energy as function of the
width of the Higgs field $w_H$ for a large fermion mass,
$m_f=1.5{\rm TeV}$ and two values of the number of populated levels.}
\end{figure}

In figure~\ref{fig:emf} we show the resulting binding energy
for our choice $m_f=1.5{\rm TeV}$. As expected we do get a bound
object for the physical value $N_C=3$ and a large number of
populated fermion levels.
It is amusing to see that even a fermion number zero object ($N_f=0$)
acquires binding. That is, for such large a fermion mass the
vacuum becomes unstable already for $N_C=3$, at least in the one fermion
loop approximation. For $m_f=170{\rm GeV}$ we need to tune $N_C\sim100$
to observe the same effect. Of course, we expect that there exists an
intermediate value of the fermion mass between $170{\rm GeV}$ and 
$1.5{\rm TeV}$ for the which this vacuum instability sets in for $N_C=3$.

\section{Conclusions}

In this paper we have studied the quantum energies of static and 
localized $W$--string configurations in a $D=2+1$ dimensional gauge 
theory. This is to be understood as a precursor to a full investigation 
of quantum energies of $W/Z$--string configurations in the standard model 
in $D=3+1$. As we know from studies on QED flux tubes, the limitation to 
$D=2+1$ dimensions is a reliable truncation for string--type configurations 
when the $D=3+1$ renormalization conditions are imposed and the vacuum 
expectation value of the Higgs field is suitably scaled.

We have concentrated on the fermion contributions to the vacuum polarization 
energy as a first calculation that comprehensively compares that energy 
to the one associated with the the trivial vacuum configuration, where
the Higgs field sits at the minimum of its potential and the gauge
fields vanish.  This approach is motivated by 
the observation that there exists a fermion zero--mode for field 
configurations that correspond to the $W$--string with the charged
Higgs field being identically zero. The population of this zero
mode could yield a significant energy 
gain. The incorporation of contributions due to a single fermion mode 
requires us to also include the full fermion vacuum polarization energy 
because it is of identical order in both the loop and large $N_C$ expansions, 
where $N_C$ counts the degeneracy of the fermions, {\it e.g.\@} the number of
color degrees of freedom. We compute the effective energy as the sum of 
the classical energy, the bound state contributions and the (renormalized) 
fermion vacuum polarization energy. In the combined limit, $\hbar\to0$ 
and $N_C\to\infty$ this approximation becomes exact at next to leading order.

We have employed techniques that use scattering data for the 
computation of the vacuum polarization energy. These techniques have 
the important and outstanding feature that standard renormalization 
conditions, formulated in terms of Green's functions in  
momentum space, can be straightforwardly imposed. This is essential for 
a sensible comparison to the energy of the translationally invariant 
vacuum configuration and to generalize the $D=2+1$ results to $D=3+1$.
To make efficient use of this powerful scattering theory approach,
however, we have to assume that the fermions in the doublet are
degenerate in order to obtain a partial wave expansion.
Our numerical results show that this contribution to the total
energy is small for configurations of interest, and does not destroy
stability of the string.

We have found energetically favored string configurations
within  an ansatz of variational parameters characterizing the
$W$--string  configuration.  Although the true minimum likely lies
outside our ansatz, its energy can only be less.  Though we find that
the fermion vacuum polarization energy reduces the effective energy, our
numerical results show that the large classical energy carried by the
pure $W$--configuration cannot easily be compensated  by fermion
quantum contributions with smaller classical energies.  Rather, in the
search for energetically stable configurations we are  led to Higgs
field dominated scenarios.  In turn, this result reduces the
importance of the zero mode.  This is a significant result, one that
runs contrary to our expectations and demonstrates that the search for
stable configurations requires one to fully account for the fermion
spectrum  in the string background, rather than merely concentrating
on just the zero mode.   Furthermore, the change in
character of the fields when going from the gauge field dominated to 
the Higgs field dominated scenarios is quite interesting.  In the gauge 
field dominated case the fields have non--zero winding at infinity 
and magnetic flux.  In the other case we are essentially left with only 
a shallow ring in coordinate space in which a hole is dug into the Higgs 
vacuum expectation value. 

In the regime of interest, where the extension of 
the background fields is a few times the Compton wave--length of 
the fluctuating fermion, the number of bound states essentially grows 
quadratically with the widths of the background fields. The hole in 
the Higgs vacuum expectation value causes the corresponding binding. 
Although these states are generally only weakly bound, 
they are so numerous that they can cause an object with large fermion
number to indeed be bound.  Though we did not observe such a
configuration for empirically motived model parameters, we have seen
that a doubling of the fermion degeneracy or an increase of the
fermion mass provides sufficient emphasis on the fermion contribution
to the total energy such that stable objects  emerge.  Typically these
stable objects carry fermion numbers of about one hundred or even more
and the configurations are quite wide, a few times the Compton
wave--length of the free fermion.

We have not taken the effort to determine a minimal value for the
fermion mass for which a stable configuration emerges. However, for 
the empirical value of the fermion mass the critical value $N_C=7$ 
is not too far away from the physical datum. It thus does not seem
totally absurd to imagine that in $D=3+1$ a stable object exists
when adopting the standard model parameters.  This certainly motivates
the corresponding extension of the present study, but some 
technical obstacles must be overcome first. In $D=2+1$ it was 
sufficient to compute Feynman diagrams (and Born terms) up to second 
order in the external fields. The expansion with respect to external 
fields is gauge variant, but only gauge invariant combinations of Higgs and 
vector fields have well defined Fourier transforms that enter the
Feynman diagrams. (The full effective energy is, of course, gauge 
invariant as the analogous peculiarity is contained in the Born series for 
the scattering data.)  Already in the present case we had to introduce 
a fake Higgs field to circumvent that problem.  Once we go to higher 
order in the Feynman diagrams, additional manipulations of this kind
will be needed.  One possibility is to unwind the string configuration at 
some distant point in space by making the angle $\xi_1$ 
in the parameterization, eq.~(\ref{eq:SphaleronSquare}) space dependent,
such that it vanishes at $r\to\infty$. Unlike in the case of the 
QED flux tubes the {\it return flux} associated with 'unwinding'
the strings will contribute to the energy even if the distant point is 
sent to infinity.  Though that contribution is easy to estimate (and 
thus subtract) for the classical energy, it is not so for the vacuum 
polarization energy, which is a non--local functional of the background 
fields.  We expect, however, that the artificial 'unwinding' part of the 
effective energy can be quantified by some approximate techniques such 
as the gradient expansion. Since the present study indicates that 
$\xi_1\to0$ for the energetically favored configuration, it is also 
worthwhile to explore pure Higgs background configurations in $D=3+1$. 
Studies in that direction are in progress.

\section*{Acknowledgments}

We are very grateful to V. Khemani for helpful cooperation in the
early stage of this investigation. O.~S. is grateful to the 
Center of Theoretical Physics at the MIT for hospitality and
the Particle Theory Group at the University of Plymouth for a
good working atmosphere and interesting discussions.

N.~G. was supported by a Cottrell College Science Award from Research
Corporation and by National Science Foundation award PHY-0555338. 
O.~S. was supported by the \textit{Deutsche Forschungsgemeinschaft} under 
grant DFG Schr 749/1--1 and by PPARC.

\appendix

\section{Perturbation Theory}
\subsection{Renormalization Conditions}
The counterterm Lagrangian for the $SU(2)_L$ model is given by
\begin{eqnarray}
{\mathcal L}_H^{\rm (ct)} & = &
c_1\tr\left(W^{\mu\nu}W_{\mu\nu}\right) +
c_2\left(\left[D^{\mu}\phi\right]^{\dag}D_{\mu}\phi\right )
+
c_3\left[\left(\phi^{\dag}\phi\right)-v^2 \right] +
c_4\left[\left(\phi^{\dag}\phi\right)-v^2\right]^2 \, .
\end{eqnarray}
The counterterm coefficients
corresponding to the renormalization conditions outlined in the main text are:
\be
c_1=\frac{g^2}{32 \pi m_f}I_1^{(3)} \,,\quad
c_2=\frac{f^2}{\pi m_f} I_2^{(3)} (\xi_H) \,,\quad
c_3=\frac{1}{\pi} f^2 m_f \quad {\rm and}\quad
\bar{c}_4 = \frac{2 f^2}{\pi} m_f I^{(3)}_4(\xi_H) \,,
\label{eq:countcoef}
\ee
with $\bar{c}_4 = 4 v^2 c_4 + c_3 + m_H^2 c_2$ and $\xi_H = m_H/m_f$.
For all parts of the Feynman diagram calculation we employ dimensional
regularization. This is the reason for $c_3$ being finite, even though
it is used to cancel a linear (by power counting) divergence. \\
The Feynman parameter integrals are given in D=2+1 by:
\bea
I_1^{(3)}(\xi)
&=&\frac{\xi+(1-\xi^2/4)\ln\frac{2+\xi}{2-\xi}}{\xi(4-\xi^2)},\\
\bar{I}_1^{(3)}(\xi) &=& 1 + \frac{\xi}{4} \ln\frac{2-\xi}{2+\xi}, \\
I_2^{(3)}(\xi) &=& \frac{1}{8 \xi^3}(4 \xi -(4+\xi^2)
\ln\frac{2-\xi}{2+\xi}), \\
I_3^{(3)}(\xi) &=& \frac{1}{8 \xi^2}(4 \xi + (4-\xi^2)
\ln\frac{2-\xi}{2+\xi}).
 \eea

\subsection{The Renormalized Second Order Feynman Diagram}

The second order renormalized Feynman diagram involves only the scalar
profile functions. It is most easily expressed using the Fourier
transforms
\be
\tilde{f}_H(k) = \int_0^{\infty} \rho d\rho J_0(k \rho) (1-f_H(\rho))
\ \textrm{and}
\ \tilde{f}_P(k) = \int_0^{\infty} \rho d\rho J_0(k \rho) f_P(\rho).
\ee
As discussed at the end of section~\ref{sec:vpe}
it suffices to consider $\xi_1=0$.
Then, the contribution to the energy is
\be
E^{(2)} = 2 m^5 \int_0^{\infty} \eta d \eta \big\{ \big[(\eta^2
  +\xi_H^2) I_2(\xi_H) + 2 I_4(\imath \eta) -2I_4(\xi_H) \big]
\tilde{f}_H^2(m_f \eta) + \big[\eta^2 I_2(\xi_H) +  I_4(\imath \eta)-1 \big]
\tilde{f}_P^2(m_f \eta)  \big\}
\ee
with $\xi_H = m_H/m_f$ and $i\eta$ being a space like momentum fraction.

\subsection{The counterterm contribution to the energy}
In this section, we give the contribution of the counterterms to the
energy that weren't used in renormalizing the second order Feynman
diagram.
The energy contribution is
\be
E_{c.t.}^{(3,4)} = 2 \pi \int \rho d\rho \, \mathcal{E}_{c.t.}
\ee
with
\be
\mathcal{E}_{c.t.} = -4c_1 \sin^2(\xi_1^2) \frac{n^2}{g^2}
\frac{f_G^{'2}}{\rho^2} + c_2 v^2 \frac{n^2}{\rho^2} \sin^2{\xi_1}
(f_H^2(1-f_G)^2 + f_P^2f_G^2)-c_4 v^4((1-f_H^2-f_P^2)^2-4(1-f_H)^2).
\ee

\newcommand{\noopsort}[1]{} \newcommand{\printfirst}[2]{#1}
  \newcommand{\singleletter}[1]{#1} \newcommand{\switchargs}[2]{#2#1}

\end{document}